\documentstyle[aps,epsfig,preprint]{revtex}
\tightenlines
\begin{document}
\draft
\title{Evolution of Fermi Liquid Behavior with Doping in the
Hubbard Model}

\author{Jungsoo Kim and D. Coffey}
\address{Department of Physics, State University of New York, 
Buffalo, NY 14260}

\date{\today}
\maketitle

\begin{abstract}
We calculate the single-particle Green's function for the
tight-binding band structure, $\xi_{\vec p}=-2t\cos p_x-2t\cos p_y
-\mu$, with a function of chemical potential $\mu$ for square-lattice
system. The form of the single-particle self-energy, $\Sigma({\vec p},
E)$, is determined by the density-density correlation function,
$\chi({\vec q}, \omega)$, which develops two peaks for $\mu \gtrsim
-2.5t$ unlike parabolic band case. Near half filling $\chi({\vec q},
\omega)$ becomes independent of $\omega$, one dimensional behavior, at
intermediate values of $\omega$ which leads to one dimensional
behavior in $\Sigma({\vec p},E)$. However $\mu \leq -0.1t$ there is no
influence on the Fermi Liquid dependences from SDW instability. The
strong $\vec p$ and $E$ dependence of the off-shell self-energy,
$\Sigma(p,E)$, found earlier for the parabolic band is recovered for
$\mu \lesssim -t$ but deviations from this develop for $\mu \gtrsim
-0.1t$. The resonance peak width of the spectral function, $A({\vec
p}, E)$ has linear dependence in $\xi_{\vec p}$ due to the $E$
dependence of the imaginary part of $\Sigma({\vec p}, E)$. We point
out that an accurate detailed form for $\Sigma({\vec p},E)$ would be
very difficult to recover from ARPES data for the spectral density.
\end{abstract}
\pacs{71.10.-w}
\section{Introduction}
The Fermi liquid(FL) theory has been used to describe metallic phase
in which quasi-particle concept develops from a free electron model to
study the metallic behavior\cite{pines}. Since the discovery of high
temperature superconductivity in the cuprates\cite{bednorz}, the
normal state properties of the quasi-two dimensional cuprates have
been investigated experimentally and theoretically. The experimental
observations such as optical conductivity\cite{timusk}, electrical
resistivity\cite{takagi,ando1,ando2} and angle resolved photoemission
spectroscopy(ARPES)\cite{schabel,norman} do not follow the
conventional FL behaviors. The absence of those FL dependences has
lead to a proposal that FL model for these quasi-low dimensional
conductors should be discarded in favor of a Luttinger liquid-like
model\cite{anderson} in which there is a separation of charge and spin
degrees of freedom in the elementary excitation.

However Castellani, Di Castro and Metzner have shown that the FL
regime is recovered for dimensions greater than one, having
investigated instability of Luttinger liquid(LL) by analyzing
correlated fermions with anisotropic hopping amplitudes in one
dimensional system\cite{castellani1} and dimensional crossover FL to
LL using analytic continuation for non-integer
dimensions\cite{castellani2}. Independent calculations show that
arbitrarily small transverse hopping kills off
LL\cite{boies,kopietz}. They also found no breakdown of the
perturbation theory.

Given that the expected FL dependences are based on parabolic band (a
spherical Fermi surface model) and a quasi-particle approximation and
assuming that the ground state is analytically continuous to the non
interacting ground state as a function of interaction strength, the
discrepancy between the FL picture and the experiment has two
possibilities. Firstly the conventional FL dependences are present but
they are limited to anomalously low energy and temperature scales
induced by interactions among the fermions or by the band structure of
the non-interacting fermions. In this picture the spectral density of
the single-particle Green's function is still characterized by a
single-particle like sharp resonance for momenta very close to the
Fermi surface whose width becomes progressively broader as the
quasi-particle momenta move further from the Fermi surface where the
quasiparticle approximation breaks down. The second possibility can be
viewed as an extreme case of the first possibility in which
interactions are so strong that the quasi-particle approximation
breaks down over at least some region of the Fermi surface. In this
case the ground state is still a FL in that it has developed
analytically out of the non-interacting ground state. This picture is
supported by ARPES measurements\cite{ding,loeser} on the underdoped
cuprates although the quasi-particle peaks remain anomalously broad
ever at the chemical potential\cite{mesot}. In these experiments the
Fermi surface seems to develop around a particular direction, signaled
by a quasi-particle peak in the measured spectral density at low
doping. The region in which there is a Fermi surface defined by the
vanishing of a quasi-particle peak grows with doping until the whole
surface is established.\cite{norman} The absence of a quasiparticle
peak is referred to as a pseudogap. Previously this type of data for
underdoped cuprates had been described in terms of hole pockets which
were shown to arise from strong correlation effects in calculations of
the t-J model\cite{trugman,moreo,eder,dagotto}. However a search for
these pockets in the recent ARPES results\cite{norman} was
unsuccessful.

The leading 2D FL behavior for on-shell self-energy with an isotropic
band structure at zero temperature, $\xi_{\vec p}={{\vec p}^2\over
2m}-{p_f^2\over 2m}$, is
\begin{mathletters}
\begin{equation}
\Sigma'({\vec p},\xi_{\vec p})=-\alpha\xi_{\vec p}-{\pi\over2}\beta
\xi_{\vec p}|\xi_{\vec p}|-\gamma|\xi_{\vec p}|^{5/2}+O (\xi_{\vec
p}^3)
\end{equation}
\begin{equation}
\Sigma''({\vec p},\xi_{\vec p})=\beta \xi_{\vec p}^2\ln |\xi_{\vec
p}/\xi_0|+O(\xi_{\vec p}^4)
\end{equation}
\end{mathletters}
where the cut-off energy $\xi_0$ is isotropic\cite{kim1}. These
leading on-shell dependences of imaginary part of self-energy had been
determined previously by a number of authors\cite{hodges,bloom}. For
low-density system $\xi_0$ goes to zero so that the region of the
generic Fermi liquid behavior is limited. The $\xi_{\vec p}|\xi_{\vec
p}|$ energy dependence in the real part is a mirror of the imaginary
part of the self-energy, $\beta\xi_{\vec p}^2\ln\xi_{\vec p}$, through
Kramers-Kronig relation which gives a $T^2$ contribution to the
specific heat in addition to that from zero sound\cite{coffey}. The
$|\xi_{\vec p}|^{5/2}$ term is the leading zero-sound
contribution. This term has a corresponding $(\xi_{\vec
p}-\xi_{th})^{3/2}$ term in $\Sigma''({\vec p},\xi_{\vec p})$, where
$\xi_{th}=(v_{zs}-v_f)p_f$, determined by the velocity of the
zero-sound mode. 2D calculation shows no evidence of a breakdown in FL
theory and is similar to 3D FL considering a parabolic band. The
cut-off energy, $\xi_0$, has a corresponding cut-off temperature,
$T_0$, which restricts the leading correction to FL behavior in
$\Sigma''(p, \xi_p, T)$ so that $T$ behavior will take over the FL
behavior at low density limit\cite{kim1}. These cut-offs are
determined by quasiparticle interactions and the characteristic energy
of the non-interacting system, $E_f$. As the system goes to quasi 1D
band which can be modeled by considering different ratio of $t_x$ and
$t_y$, for example quasi 1D organic materials such as $(TMTSF)_2X$ or
$BEDT-TTF$\cite{williams}, van Hove singularity moves to the bottom of
the band so that particle-hole($ph$) pair contribution is more
important than particle-particle($pp$) contribution for low density
limit. In this case the nature of the q1D FL behavior remains similar
to that of 2D FL\cite{kim4}. For parabolic band structure $pp$ channel
which leads Cooper instability is more important than $ph$ channel in
3D and vice versa in 1D due to the density of states. No particular
set of diagrams is significant in 2D\cite{kim2}, but at least the
second order calculation is shared by both of them so that functional
dependence in either one will represent the FL behavior. The $pp$
channel is discussed later in this paper.

The cuprates have an anisotropic band-structure which can lead to
strong band structure features in the density of states. This
anisotropy of the band structure is used in a number of different
models to explain the lack of FL dependences in
experiment\cite{virosztek,ruvalds}. The anisotropy has been invoked to
describe transport data in the hot spot\cite{hlubina} and cold
spot\cite{ioffe} models. In the hot spot model it is proximity to the
SDW instability that is important whereas in the cold spot model it is
proximity to the superconducting phase through pairing fluctuations
which leads to the anisotropy. More recently renormalization group
calculation have shown that if the Fermi surface is dominated by a set
of saddle points in the density of states Landau FL theory, the
quasi-particle approximation to FL theory, can
breakdown\cite{furukawa1,furukawa2}.

The Hubbard model has been employed extensively in theoretical
investigations of the normal state properties of the cuprates because
the tight binding band structure and short-range Coulomb correlations
result in an effective magnetic Hamiltonian for low doping which is
consistent with the data. We investigate the FL characteristics of
this model as a function of chemical potential, $\mu$, from $\mu=-3t$
where the Fermi surface is a circle to $\mu=-0.1t$ , near
half-filling, where it is almost a square. A 2D nearest neighbor
tight-binding band structure, $\xi_{\vec p}=-2t\cos p_x-2t\cos p_y
-\mu$, will be used to determine the momentum and energy dependence of
single properties in FL behavior where t is the nearest neighbor
hopping constant. Near the half-filling curved $\bar M$ points are
connected by flat Fermi surface which is a good model for the band
structure of BSCCO in which there is a strongly nested
vector\cite{dessau}. The Hamiltonian includes a repulsive contact
interaction which we treat in the weak coupling approximation. By
introducing a cut-off, $q_c$, for the interaction in momentum space
the influence of long wavelength interactions on $\Sigma({\vec p},E)$
in the $ph$ channel has been investigated in this model and shown to
determine the functional form of $\Sigma({\vec p},E)$ at low energies
as in the parabolic band case. The evolution of the leading behaviors
in real and imaginary part is determined by the changes in the bare
$ph$ propagator as the Fermi surface becomes more anisotropic.  First
we calculate the contribution to $\Sigma({\vec p},\xi_{\vec p})$, the
on-shell self-energy, to the second order in $V$ which gives the
functional form of the corrections and then consider the effect of the
repeated scattering from $ph$ channel which would indicate the
influence of any instabilities in these behaviors. 

We next calculate the off-shell self-energy, $\Sigma({\vec p},E)$,
which is probed directly in angle resolved photoemission through the
spectral density $A({\vec p}, E)$. It is found that energy dependence
of the imaginary part of $\Sigma({\vec p},E)$ leads to a width of the
quasi-particle resonance in $A({\vec p},E)$ which is linear in
$\xi_{\vec p}$ even when the on-shell $\Sigma''({\vec p},E)$ has the
expected $\xi_{\vec p}^2 \ln \xi_{\vec p}$ dependence. We point to the
difficulty of extracting the functional form of off-shell self-energy
from ARPES data on the spectral function.
\section{Calculations}
From the Hamiltonian
\begin{equation}
H=\sum_{{\vec p}, \sigma} \xi_{{\vec p}} c^\dagger_{{\vec
p},\sigma}c_{{\vec p},\sigma}+ \sum_{{\vec p},{\vec q},\sigma,
\sigma\prime}V(\vec q) c^\dagger_{{\vec p},\sigma}c_{{\vec p}
',\sigma\prime} c^\dagger_{{\vec p} ' -{\vec q},\sigma\prime}c_{{\vec
p} +{\vec q},\sigma}
\end{equation}
where $\xi_{\vec p}$ is the electronic band structure and $V$ is the
electron-electron interaction which is treated as a perturbation term,
the single-particle self-energy, $\Sigma(\vec p,E)=\Sigma'(\vec
p,E)+\imath \Sigma''(\vec p,E)$, is given by
\begin{equation}
\Sigma({\vec p}, \imath E_n)=-T\sum_{{\vec q}, 
\omega_l}G({\vec p} - {\vec q}, \imath E_n -\imath \omega_l)
V_{eff}({\vec q}, \omega_l)
\label{selfeq}
\end{equation}
where $G({\vec p}, \imath E_n)$ is the unperturbed temperature Green's
function and $\omega_l$ is Bose Matsubara frequencies. The $V_{eff}$
includes all the $ph$ diagrams in principle.  As a function of filling
the electronic band structure develops an anisotropic Fermi surface
which will end up with quasi-one dimensional square Fermi surface at
$\mu=0$.  As the system goes to half-filling, the density of states at
the Fermi surface approaches to the Van Hove singularity, which makes
$ph$ channel more significant in the low filling system. We are to
concentrate on two directions in momentum space through out the
calculations, $(1,0)p$ and $(1,1)p$, which are extreme directions in
the tight-binding band structure.

The accurate calculation of the density-density correlation function
is essential in this study. The correlation function, $\chi(\vec
q,\omega)=\chi'(\vec q,\omega)+\imath\chi''(\vec q,\omega)$, for $ph$
channel is given by
\begin{equation}
\chi(\vec q, \omega)=\sum_{\vec p}{f(\xi_{\vec p+\vec q})-f(\xi_{\vec
p})\over \omega - (\xi_{\vec p+\vec q} -\xi_{\vec p})}
\end{equation}
where $\xi_{\vec p}=-2t\cos p_x-2t\cos p_y -\mu$. The imaginary part
of the density-density correlation function for the tight-binding band
structure is given by
\begin{mathletters}
\begin{equation}
\chi''(\vec q, \omega > 0)={1\over 4\pi}\int dp_x \int dp_y
{\Theta(\xi_{\vec p})\Theta(-\xi_{\vec p-\vec q})\delta(p_y-p_y^*)
\over |-4t\sqrt{\sin^2{q_y\over2}-[{\omega\over 4t}-\sin(p_x-{q_x\over
2})\sin{q_x\over 2}]^2}|}
\end{equation}
\begin{equation}
p_y^*=\sin^{-1}[{\omega -4t\sin(p_x-{q_x\over2})\sin{q_x\over 2}\over
4t\sin{q_x\over2}}]+{q_y\over2}
\end{equation}
\end{mathletters}
in which subscripts, $x$ and $y$, can be switched when $q_y <
q_x$. The Kramers-Kronig relationship is used to calculate the real
part of the correlation function. The $(1,0)q$ or $(0,1)q$ direction
can be handled by the expression,
\begin{equation}
\chi''_{10q}(\omega)=
{1\over 2\pi}
{p_f(p_1)\Theta(p_{f,max}^2-p_1^2)-p_f(p_2)\Theta(p_{f,max}^2-p_2^2)
\over \sqrt{(4t\sin{q\over 2})^2-\omega^2}}
\Theta[1-({\omega \over 4t\sin{q\over 2}})^2]
\end{equation}
where $p_f(p)=\cos^{-1}[-{\mu\over 2t}-\cos(p)]$, $p_{f, max}=
\cos^{-1} [-1-{\mu\over 2t}]$, $p_1=\sin^{-1}[{\omega \over
4t\sin(q/2)}]-{q\over 2}$ and $p_2=\sin^{-1}[{\omega \over
4t\sin(q/2)}]+{q\over 2}$. Unlike the parabolic band structure the
anisotropy of the system in the tight-binding band structure restricts
the solution of the delta function to different areas of the Fermi
surface as the chemical potential approaches to half-filling and a
direction of the system aligns $(1,1)q$. For a given $\omega$, the
solutions of the delta function come from two different regions of
Fermi surface. Only one of these contribute for the $\vec q=(1,0)q$
direction, whereas both of them survive in $\vec q=(1,1)q$ direction
in the long wavelength limit.

On the $ph$ channel the short wavelengths do not affect the
logarithmic behavior in $\Sigma''({\vec p},\xi_{\vec p})$ but do
contribute to the cut-off energy, $\xi_0$. At half-filling with
$q=2p_f$, the correlation function diverges in both real and imaginary
part which will reflect SDW instability\cite{hirsch,lee}. Hence by
restricting a range of the interaction we tune the cut-off momentum,
$q_c$, in the self-energy calculation to identify the values of $q$
which are responsible for FL energy dependence.

First we discuss results of the 2nd order calculation in the next
section which will show the FL energy dependences in the function of
band structure and we evaluate and discuss the repeated scattering
case and Stoner instability. Except in $\mu \rightarrow 0$ limit,
there is no nesting phenomena from short wave length which solely
contribute the cut off energy, $\xi_0$ without affecting the log
behavior. We consider two different regions of $\mu$ separately as in
anisotropic 2D FL and quasi 1D FL.
\subsection{2nd order calculation of $\Sigma({\vec p},\xi_{\vec p})$}
The self-energy at zero temperature for second order calculation is
given by
\begin{equation}
\Sigma(\vec p, E > 0)=\sum_{\vec q}\Theta(E-\xi_{\vec p-\vec
q})\Theta(\xi_{\vec p-\vec q})V_{\vec q}^2\chi(\vec q,E-\xi_{\vec
p-\vec q})
\label{stepeq}
\end{equation}
where $\chi(\vec q,\omega)$ is the density-density correlation
function or response function and $V_{\vec q}=V\Theta(q_c-|\vec q|)$
in which $q_c=\pi$ and $V=t$ throughout the calculations. The
$\xi_{\vec p}^2\ln\xi_{\vec p}$ behavior in the imaginary part of the
self-energy comes from the linear energy dependence in $\chi''(\vec q,
\omega)$. In the self-energy calculation the energy dependence is
mainly determined by the frequency dependence of the density-density
correlation function rather than the step functions in the expression
for the self-energy. Since the second order contribution is common for
both of $ph$ and $pp$ channel, the second order calculation reveals
the functional dependence of FL behavior as a function of chemical
potential which we break in two regions, 2D and quasi-1D.

\subsubsection{Anisotropic 2D Fermi Liquid : $-4t\leq\mu\leq-t$}
\begin{figure}
\begin{center}
\epsfig{file=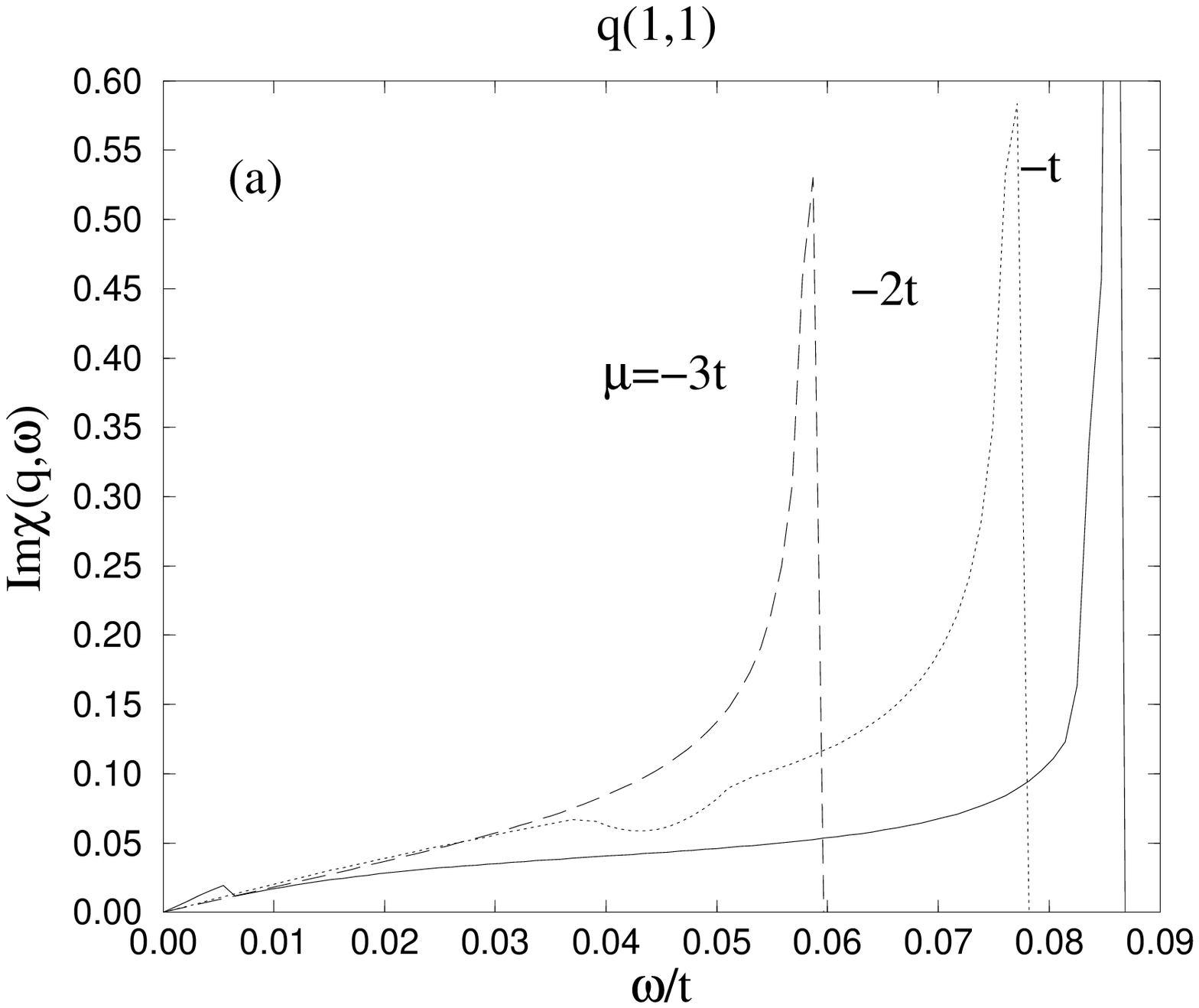,width=3.2in,height=3.5in}
\epsfig{file=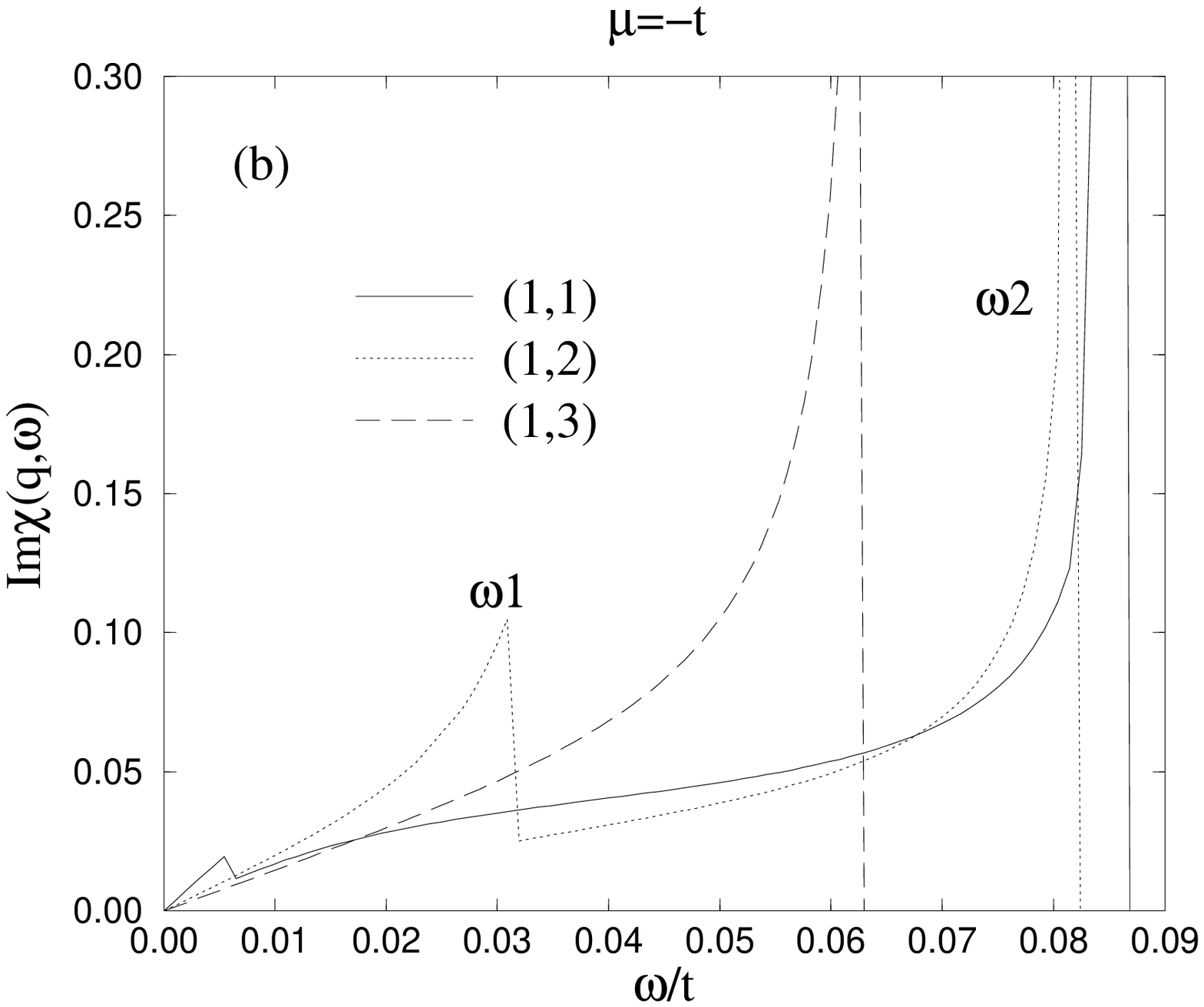,width=3.2in,height=3.5in}
\epsfig{file=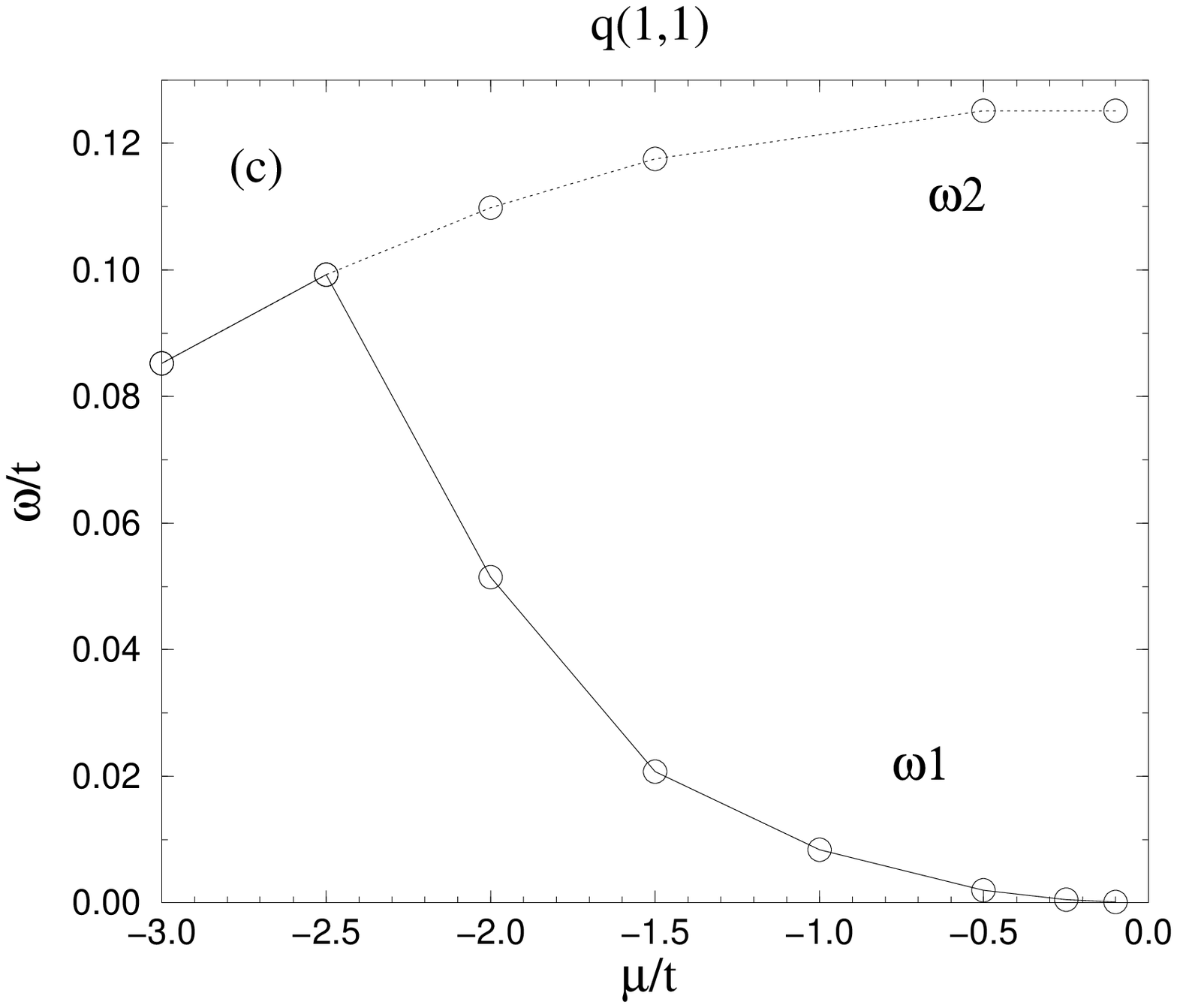,width=3.2in,height=3.5in}
\end{center}
\caption{The correlation functions as a function of $\mu$ and
directions for $|{\vec q}|=0.01\pi$. (a) $\chi''$ as a function of
$\mu$. Two peaks are appeared in $\chi''$ for the anisotropic band
structure. (b) $\chi''$ as a function of directions in $\vec q$. As
$\vec q$ points $(1,0){\vec q}$, $\omega_1$ gets closed to
$\omega_2$. (c) The locations of peaks in the correlation function,
$\omega_1$ and $\omega_2$. As $\mu \rightarrow 0$ the location of the
first peak approaches to zero.}
\label{chi2d}
\end{figure}

As the chemical potential, $\mu$, goes to zero, flatness of Fermi
surface along $(1,1)p$ direction develops. Although the Fermi surface
develops flat, quasi-one dimensional, regions for $\mu \sim -t$, no
signature of quasi-one dimensional behavior is seen in $\Sigma({\vec
p},E)$. The figure \ref{chi2d} shows the evolution of the response
function with a function of $\mu$ in the directions of $(1,0)q$,
$(1,2)q$ and $(1,1)q$ with $q=10^{-2}\pi$. The cut-off energies,
$\omega_1^*$ and $\omega_2^*$, associated with the contributions given
momentum $\vec q$ are from the two different Fermi surface
contributions through the delta function for a given momentum $\vec
q$. Those contributions give two different locations of the divergent
peaks, $\omega_1^*$ and $\omega_2^*$. As $\vec q$ points in the
$(1,1)$ direction, $\omega_1^*$ approaches zero frequency so that the
FL linear behavior is limited to small energies. However even in
$(1,1)q$ direction, the linear piece survives and $\Sigma({\vec p},E)$
has FL behavior for $\mu=-t$. As $\omega_1^*$ goes to zero energy, the
linear behavior in $\chi''({\vec q},\omega)$ is limited to a small
$\omega$ region above which $\chi''({\vec q}, \omega)$ is independent
of $\omega$ as $\mu \rightarrow 0$ as in one dimensional system. This
will be discussed later. Figure \ref{chi2d} shows the anisotropic band
structure effects, $\omega_1^*$ and $\omega_2^*$ feature, in
particle-hole correlation function which first appear around
$\mu=2.5t$.  Except $\mu \sim 0$ case, the $\chi''({\vec q},\omega)$
follows $\omega /|{\vec q}|$ at small $\omega$ which leads to
$\xi_{\vec p}^2\ln\xi_{\vec p}$ form for $\Sigma''({\vec p},\xi_{\vec
p})$.

\begin{figure}
\begin{center}
\epsfig{file=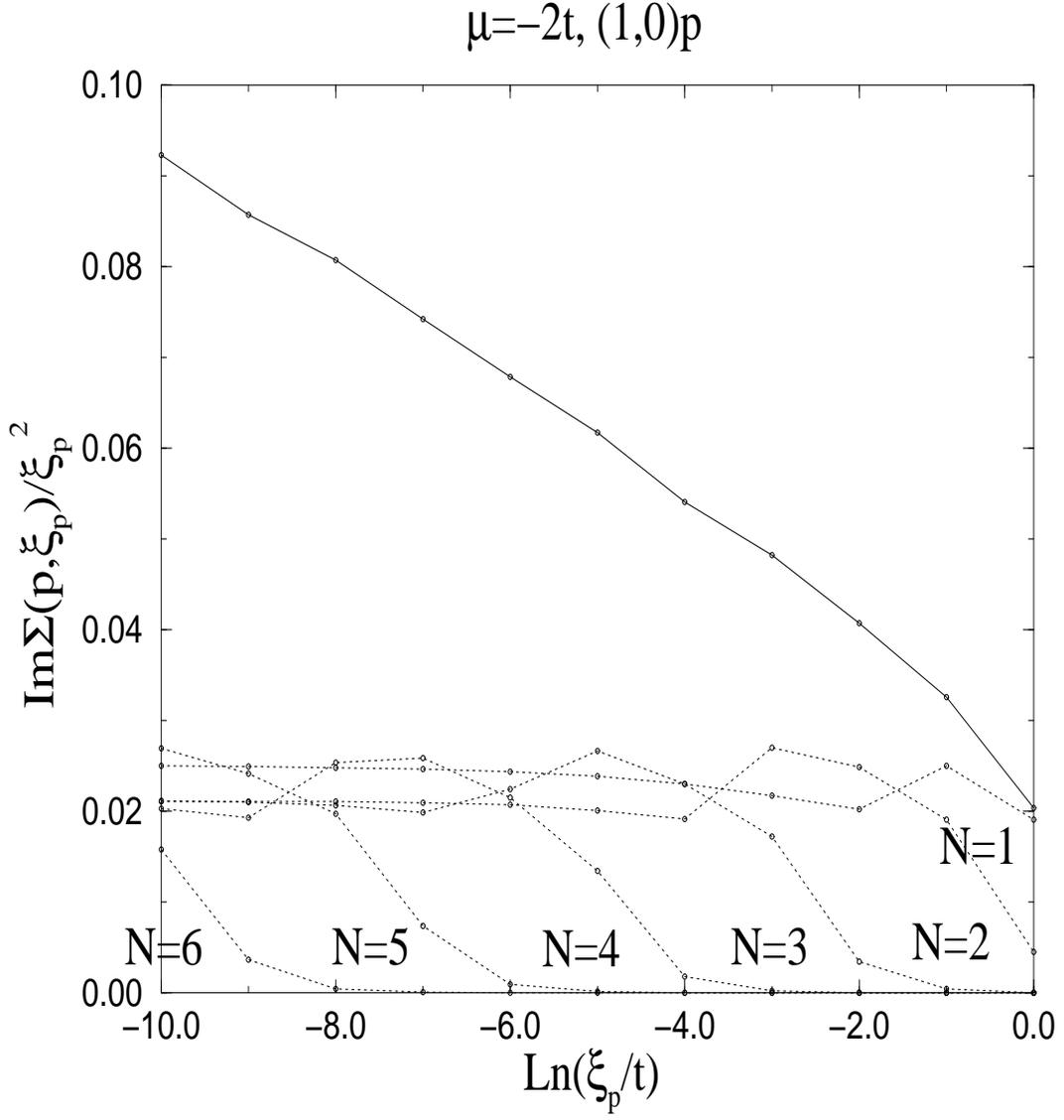,width=5.5in,height=6in}
\end{center}
\caption{The momentum contributions to the self-energy in second order
calculations with $q=10^{-N}\pi$. Each momentum is responsible for
the log-behavior to the corresponding energies.}
\label{selfdemo}
\end{figure}

The log behavior of the imaginary part of the on-shell self-energy
comes from long wave length calculation. The figure \ref{selfdemo}
demonstrates contributions $|\vec q| \leq {\pi \over 10}$ in
$\chi({\vec q},\omega)$ to the imaginary part of on-shell self-energy
with six different sets of $|\vec q|$ values given by $10^{-n}\pi \leq
|{\vec q}| \leq 10^{-(n-1)}\pi$ where $n=6, 5, 4, 3, 2, 1$. For small
values of $|\xi_{\vec p}/t|$ the contributions from longer wavelengths
provide the $\xi_{\vec p}^2 \ln \xi_{\vec p}$ dependences whereas
shorter wavelengths lead to contributions to the $\xi_{\vec p}^2$
dependence. The log behavior is restricted by small energy and the
$|{\vec q}|$s which are long wavelength. The anisotropic 2D systems
with $-4t \leq \mu \leq -t$ shows generic FL behavior with increasing
anisotropy in $\vec p$. The correction term is enhanced by tuning the
chemical potential close to zero and will be discussed in the next
section. The cut-off,$\xi_0$, and $\beta$ for the two directions of
$\vec p$ reflect the growing anisotropy of the Fermi surface as $\mu$
changes from $-3t$ to $-t$.
\subsubsection{Quasi-1D behavior : $-t\leq\mu\leq 0$}
The purely one dimensional response function with a tight-binding band
structure, $\xi_p=-2t\cos p-\mu$, is
\begin{equation}
\chi_{1D}''(q, \omega)={1\over 2}{\Theta(p_f-|{q\over
2}-\sin^{-1}({\omega \over 4t\sin{q\over 2}})|)-\Theta(p_f-|{q\over
2}+\sin^{-1}({\omega \over 4t\sin{q\over 2}})|) \over
|4t\sin{q\over2}\sqrt {1-({\omega \over 4t\sin{q\over
2}})^2}|}\Theta(1-|{\omega \over 4t\sin{q\over 2}}|^2)
\end{equation}
where $p_f=cos^{-1}(-\mu/2t)$ is the Fermi level for one dimensional
system. In the small frequency limit there is no energy dependence in
$\chi_{1D}''$ which has corresponding logarithmic divergence in
$\chi_{1D}'$,
\begin{equation}
\chi_{1D}'(q, \omega=0)={1 \over -4t\pi \sin(q/2)}\ln|{\tan(p_f-q/2)
\over \tan(p_f+q/2)}|.
\end{equation}
At static limit this divergence
at $q=2p_f$, so called perfect nesting vector, drives the Peierls
instability via Kohn anomaly which leads to the metal-insulator
transition\cite{rice}.

\begin{figure}
\begin{center}
\epsfig{file=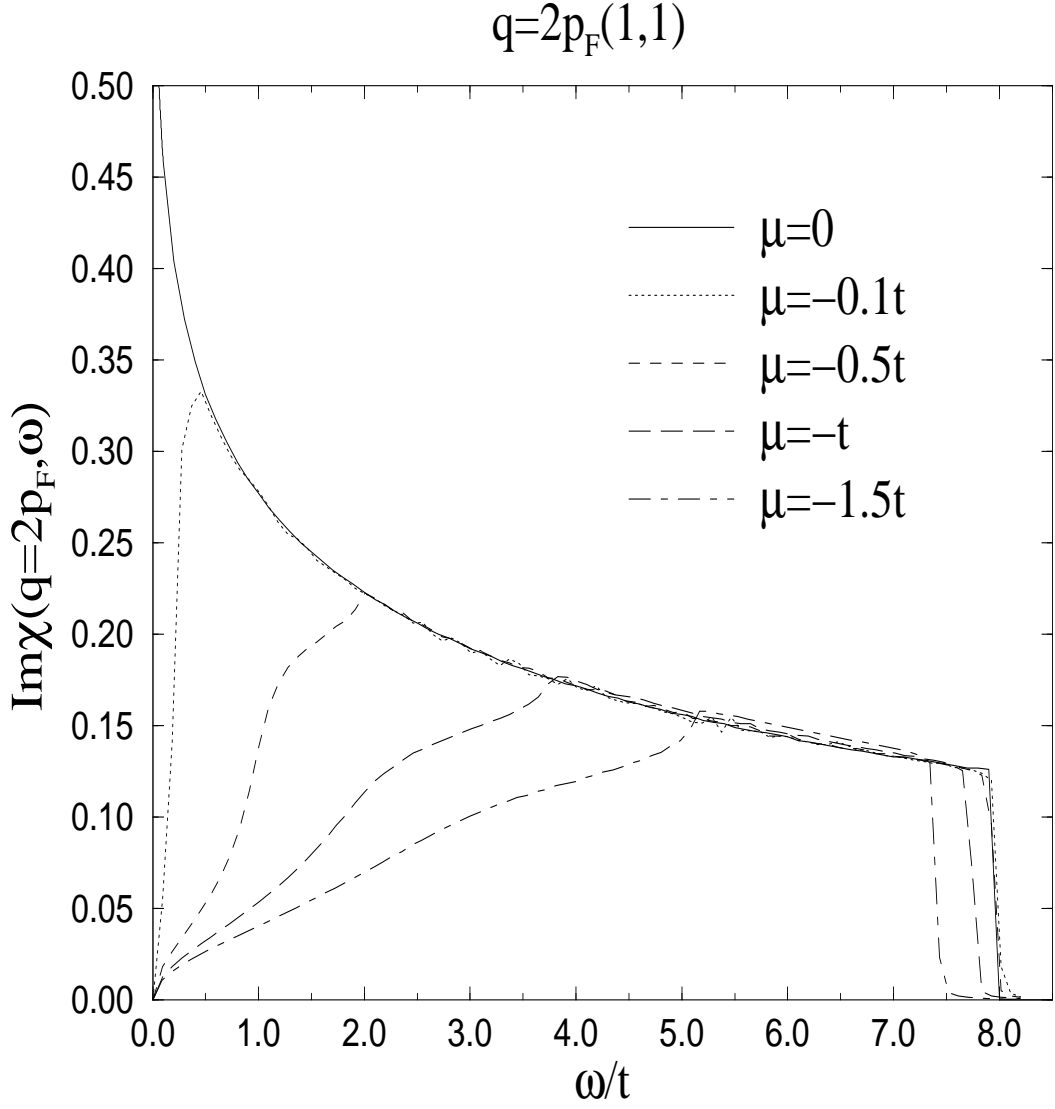,width=5.5in,height=6in}
\end{center}
\caption{The correlation functions for $q=2p_F$ with different
$\mu$s. As $\mu \rightarrow 0$ logarithmic divergence is rapidly
developed near $\mu \sim 0$.}
\label{chi2pf}
\end{figure}

Coming back to 2D problem, at $\mu=0$ the system is one dimensional in
that there is perfect nesting between parts on the Fermi surface along
$(1,1)\vec p$ direction. Nesting vector, $\vec q=(1,1)\pi$, gives a
form
\begin{equation}
\chi''(\pi,\pi,\omega ; \mu=0)= {\pi\over 2} sign(\omega) N(\omega/2)
\end{equation}
where $N(\omega)$ is density of state which has logarithmic divergence
at $\omega=0$, which breaks quasi-particle picture. For $\mu=0$ the
constant energy dependence at long wavelengths in $\chi''({\vec
q},\omega)$ which is one dimensional effect lasts up to zero
energy. Using the Kramers-Kronig relationship this energy dependence
for small $|{\vec q}|$ in $\chi''({\vec q},\omega)$ leads to a $\ln
\omega$ divergences in the real part, $\chi'({\vec q},\omega)$,
signaling the break down of RPA. The large $|\vec q|$ behavior is
shown in figure \ref{chi2pf} where it is clear that the $\ln \omega$
divergence in $\chi''({\vec q},\omega)$ is rapidly destroyed as $\mu$
goes away from half-filling.

\begin{figure}
\begin{center}
\epsfig{file=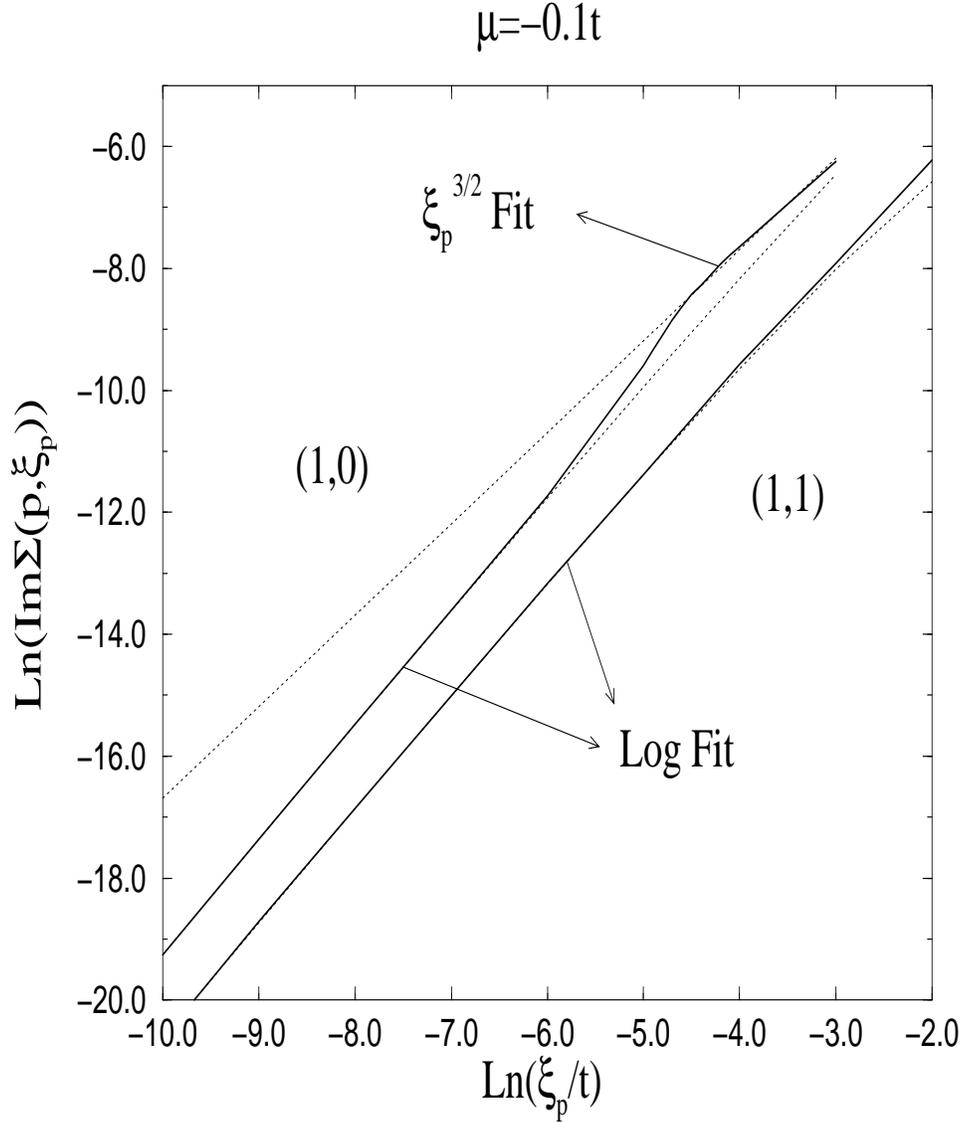,width=5in,height=6in}
\end{center}
\caption{The imaginary part of self-energy in second order
calculations with $V=t$ shows $\xi^{3/2}$ behavior which is coming
from frequency independent $\chi''$. This crossover from $\xi_{\vec
p}^2 \ln \xi_{\vec p}$ to $\xi_{\vec p}^{3/2}$ appears in $(1,0)p$
direction. The dotted lines are fitted lines.}
\label{se3_2}
\end{figure}

For $|\vec q|<10^{-1}\pi$ structure of $\chi''$, two divergent peaks
identified by $\omega_1^*$ and $\omega_2^*$, follows the discussion in
previous section. Due to rapid development of frequency independence
in $\chi''$ for low energies as $\mu \rightarrow 0$, the quasi 1D
corrections to $\xi_{\vec p}^2 \ln \xi_{\vec p}$ behavior of
$\Sigma''({\vec p},\xi_{\vec p})$ at intermediate values of $\omega$
have a $\xi_{\vec p}^{3/2}$ dependence in the quasi-particle life
time. This is shown in figure \ref{se3_2} for $\mu=-0.1t$ in which the
Fermi surface is flat over an extended area. The $\xi_{\vec p}^{3/2}$
correction sets in at $\xi_{\vec p} \sim 0.02t$, which limits generic
FL behavior to small region. As $\mu$ goes away from half filling, the
logarithmic divergence in $\omega$ dies out rapidly as in the figure
\ref{chi2pf} so that the system is far from a Peierls-like
instability. The second order calculation shows that quasi-particle
life time has generic FL dependence with $\mu \neq 0$ with a
$\xi_{\vec p}^{3/2}$ term as a correction to $\xi_{\vec p}^2 \ln
\xi_{\vec p}$ when $\mu \lesssim 0$.

\subsection{RPA calculation of $\Sigma({\vec p},\xi_{\vec p})$}
\subsubsection{Long wavelength contribution}
\begin{figure}
\begin{center}
\epsfig{file=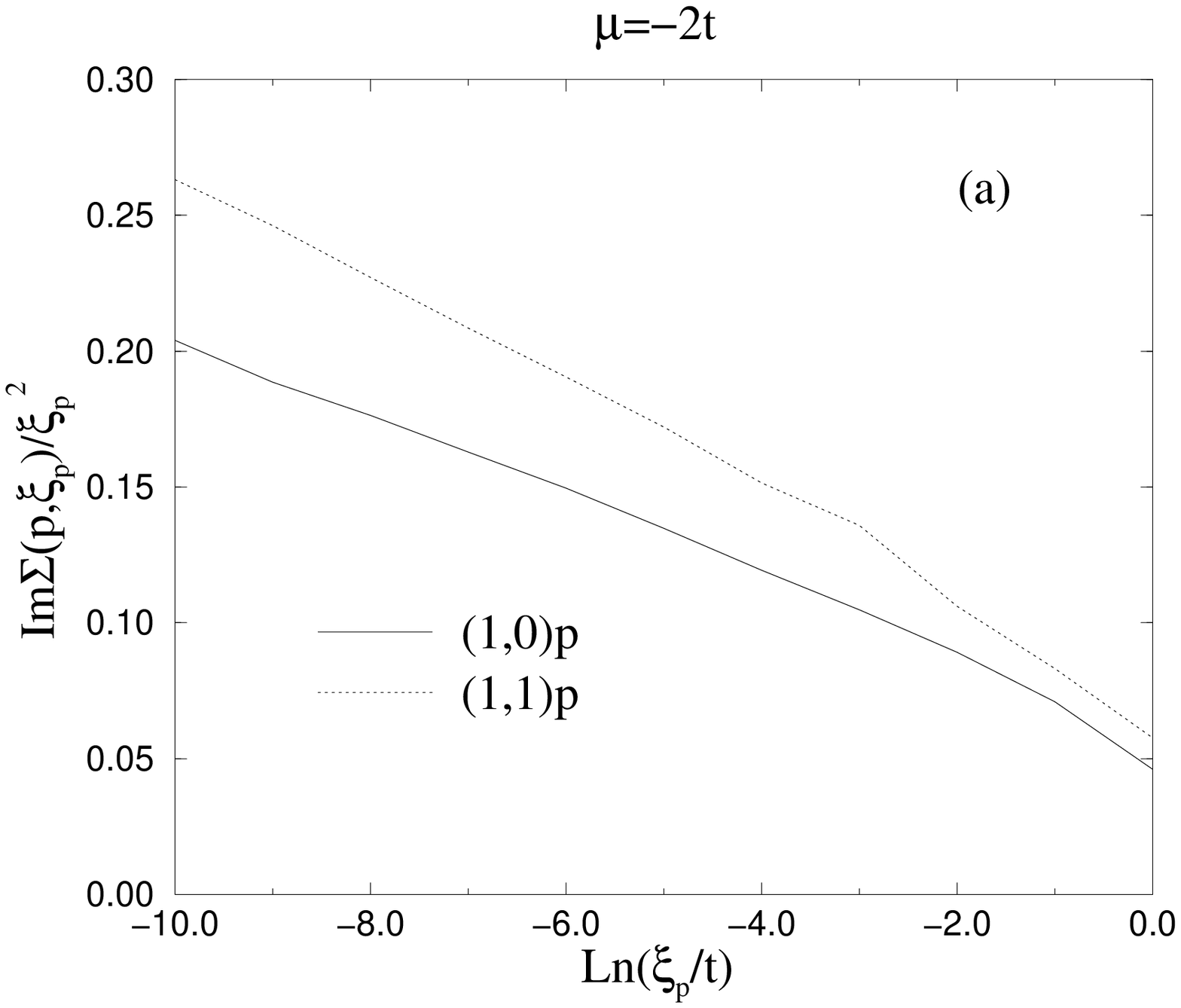,width=5in,height=4in}
\epsfig{file=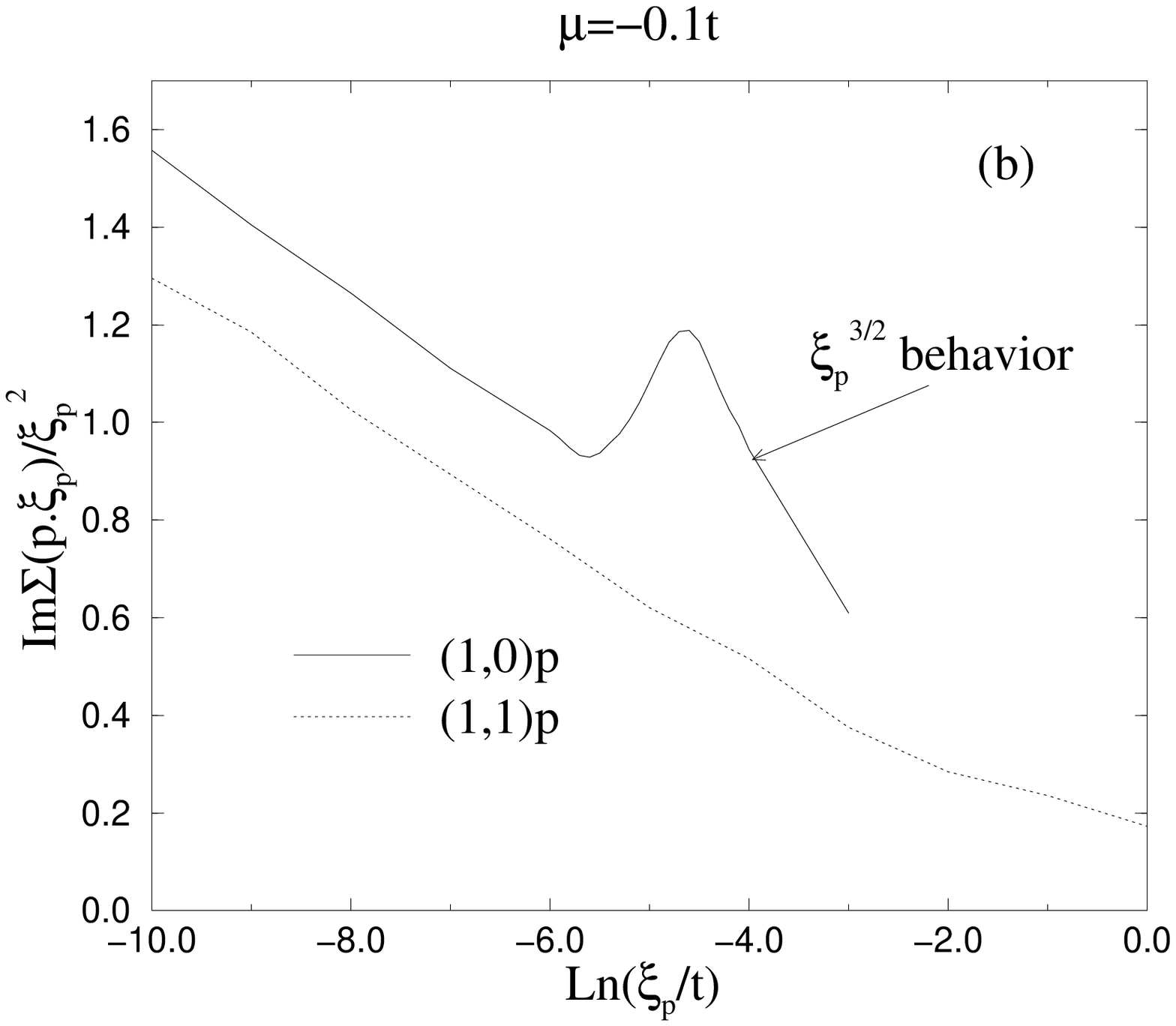,width=5in,height=4in}
\end{center}
\caption{The imaginary part of self-energies in RPA with $V=t$ for
$(1,0)p$ and $(1,1)p$ directions with (a) $\mu=-2t$ and (b)
$\mu=-0.1t$. For $\mu=-0.1t$ case, $\xi_{\vec p}^{3/2}$ behavior for
$(1,0)p$ direction appears as in 2nd order calculation. Both of the
directions show the logarithmic behavior.}
\label{seirpa}
\end{figure}

\begin{figure}
\begin{center}
\epsfig{file=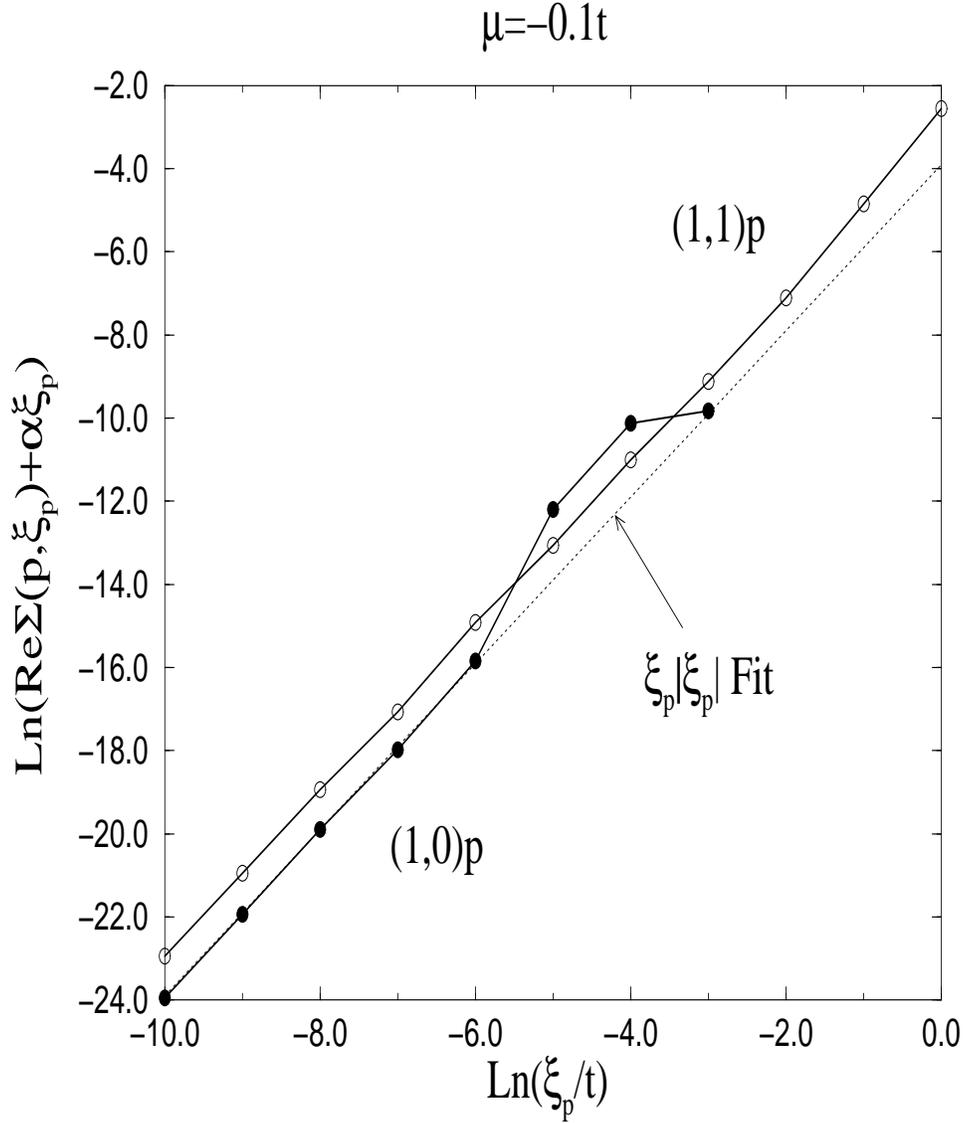,width=5in,height=6in}
\end{center}
\caption{The real part of self-energy calculated in RPA with $V=t$
shows $\xi|\xi_{\vec p}|$ behavior correction to $\alpha \xi_{\vec p}$
term. The slope 2 indicates the power of the fit. The deviation from
the $\xi_{\vec p}|\xi_{\vec p}|$ behavior comes from the long
wavelenght quasi 1D frequency dependence of $\chi({\vec q},\omega)$.}
\label{serrpa}
\end{figure}

With $\mu \neq 0$ the long wavelength limit contribution which is
responsible for $\xi_{\vec p}^2 \ln \xi_{\vec p}$ or $\xi_{\vec
p}^{3/2}$ for $\mu \sim 0$ survives unless the strong interaction
limit is considered in RPA. The effective interaction in equation (2)
for repeated scattering involves in two independent channels, the
symmetric and the antisymmetric channels, followed by spin exchanges
of 0 and 1 and is given by
\begin{equation}
{V}_{eff}({{\vec q}},\omega)={1\over 2}{V^2_s\chi
({\vec q}, \omega)\over {1-V_s\chi ({\vec q}, \omega)}}+
{3\over 2}{V^2_a\chi
({\vec q}, \omega)\over {1-V_a\chi ({\vec q}, \omega)}}
\end{equation}
where $V_s=V_q$ and $V_a=-V_q$. In RPA the imaginary part of
self-energy, $\Sigma''({\vec p},\xi_{\vec p})$, is just enhanced from
the second order calculations as in figure \ref{seirpa}. $\mu=-0.1t$
case includes $\xi_{\vec p}^{3/2}$ corrections in $(1,0)p$
direction. The real part, $\Sigma'({\vec p},\xi_{\vec p})$, has
$\xi_{\vec p} |\xi_{\vec p}|$ correction to $\xi_{\vec p}$ term. In
the figure \ref{serrpa} the $\xi_{\vec p}$ term is extracted and comes
from long wavelength limit. The $\xi_{\vec p} |\xi_{\vec p}|$ term is
a mirror of $\xi_{\vec p}^2 \ln \xi_{\vec p}$ term in $\Sigma''({\vec
p},\xi_{\vec p})$ through Kramers-Kronig relationship and the
deviations from the fits are consistent to both of $\Sigma''({\vec
p},\xi_{\vec p})$ and $\Sigma'({\vec p},\xi_{\vec p})$ graphs. The
on-shell calculation of $\xi_{\vec p}|\xi_{\vec p}|$ term in (1,0)
direction with $\mu=-0.1t$ stops at $\xi_{\vec p} \sim 0.05t$ because
the phase space runs out for that direction and the correction to the
$\xi_{\vec p}|\xi_{\vec p}|$ term sets in earlier than $(1,1)$
direction. The coefficients of the $\xi_{\vec p}|\xi_{\vec p}|$ terms
for the two directions are different due to anisotropy of the system.

\begin{figure}
\begin{center}
\epsfig{file=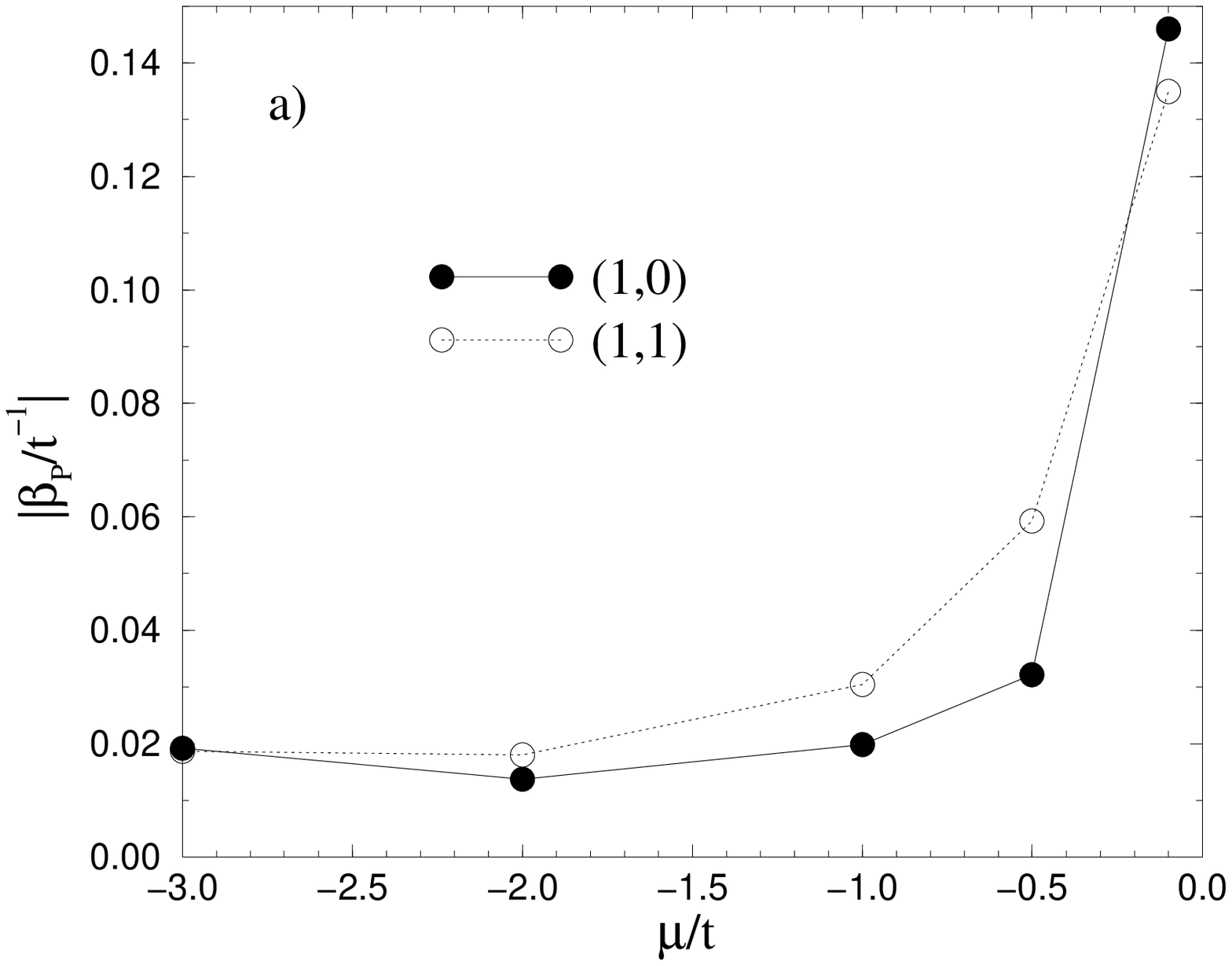,width=5in,height=4in}
\epsfig{file=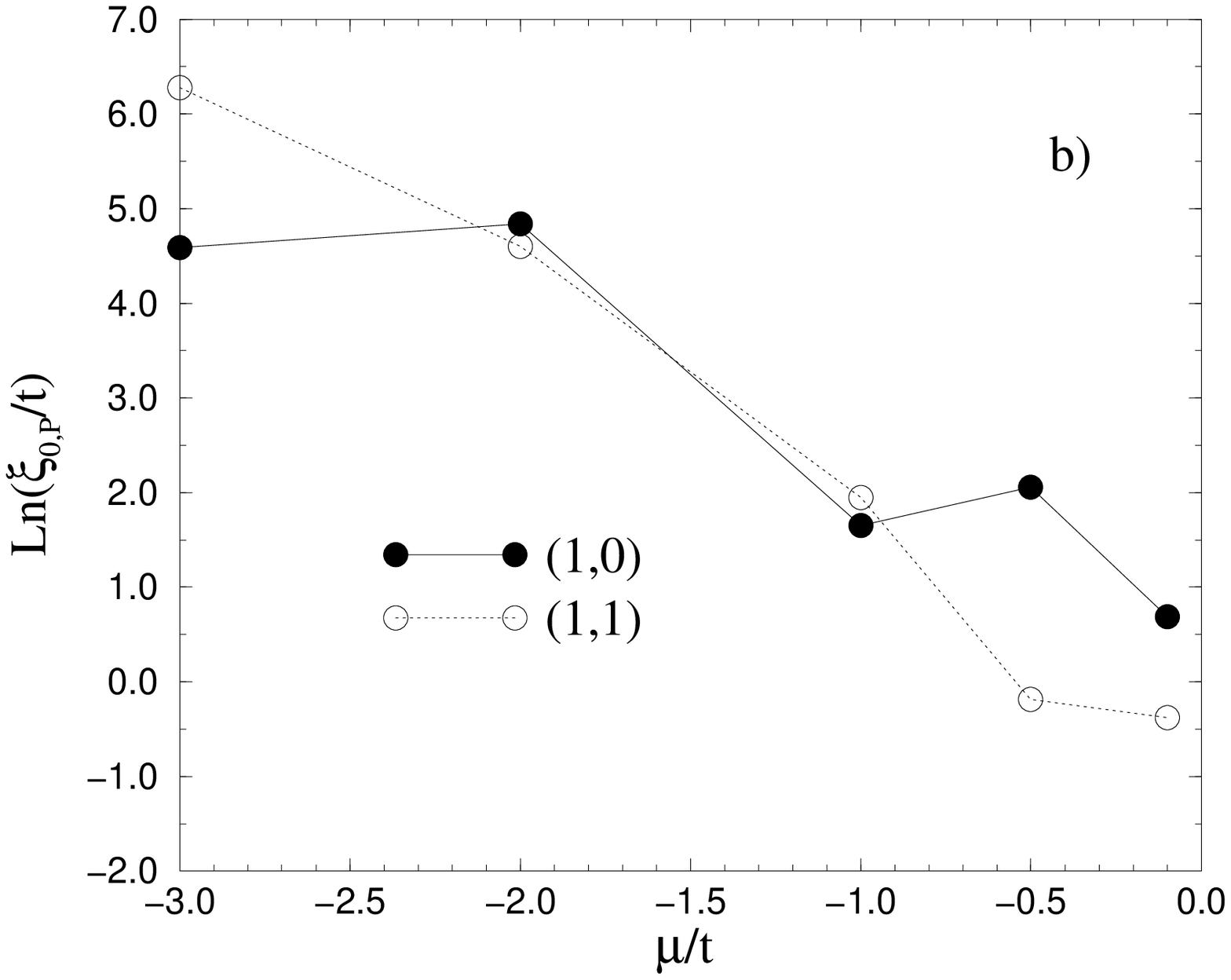,width=5in,height=4in}
\end{center}
\caption{The slopes, $\beta_{\vec p}$, (b) and the cut-off energies,
$\xi_{0,{\vec p}}$ (a) in RPA as a function of $\mu$ from the form,
$\Sigma''({\vec p},\xi_{\vec p})=-\beta_{\vec p} \xi_{\vec p}^2\ln
{\xi_{\vec p}\over \xi_{0,{\vec p}}}$.}
\label{cutoff}
\end{figure}

The figure \ref{cutoff} shows the functional dependence of the cut-off
energy in imaginary part of the self-energy, $\xi_{0,{\vec p}}$, and
the coefficients, $\beta_{\vec p}$, for the tight-binding band
structure. The anisotropy from $\beta_{\vec p}$ grows as the band is
filled, but near half filling the trend changes rapidly due to the
flatness of the band structure. The cut-off energy shows anisotropy
even at $\mu=-3t$ which is different from the figure \ref{chi2d} in
which the anisotropic effects of splitting the two different peaks in
$\chi''({\vec q}.\omega)$.
\subsubsection{Instabilities}
As discussed above the correlation function itself does not indicate
any instability unless $\mu=0$ as in figure \ref{chi2pf}. For the
repulsive interaction antisymmetric channel in effective interaction
could have an instability indicated by a pole at $\omega=0$ in the
effective interaction. The solution for it is given by $1/V_{\vec
q}=\chi'({\vec q},\omega=0)$ in RPA in which there is divergence with
perfect nesting vector at $\mu=0$ so that there are solutions for all
$|V_{\vec q}|$'s. Within the RPA the question is how fast the system
goes to 1D as a band approaches to half-filling.

In parabolic band structure, the real part of the correlation function
in static limit is given by
\begin{mathletters}
\begin{equation}
\chi_{1D}(q)=N_{1D}(0){1\over 2(q/2p_f)}\ln|{1+(q/2p_f)\over
1-(q/2p_f)}|
\end{equation}
\begin{equation}
\chi_{2D}(q)=N_{2D}(0)[1-{\sqrt {(q/2p_f)^2-1}\over
(q/2p_f)}\theta\{(q/2p_f)-1\}]
\end{equation}
\begin{equation}
\chi_{3D}(q)=N_{3D}(0)[{1\over 2}+{1-(q/2p_f)^2 \over
4(q/2p_f)}\ln|{1+(q/2p_f)\over 1-(q/2p_f)}|]
\end{equation}
\end{mathletters}
where $N(0)$s are the densities of state on Fermi level for each
dimensions and $\theta$ is a step function. In 2D the correlation
function has a cusp and a discontinuous derivative at $q=2p_f$ due to
the step function whereas 3D case varies smoothly. Since the Fermi
surface consists of two points in 1D, the nesting condition $\xi_{\vec
p}-\xi_{p+2p_f}$ is satisfied over the entire Fermi surface resulting
in a log divergence at $q=2p_f$ which is responsible for the breakdown
of the quasiparticle picture leading to a metal-insulator transition.

\begin{figure}
\begin{center}
\epsfig{file=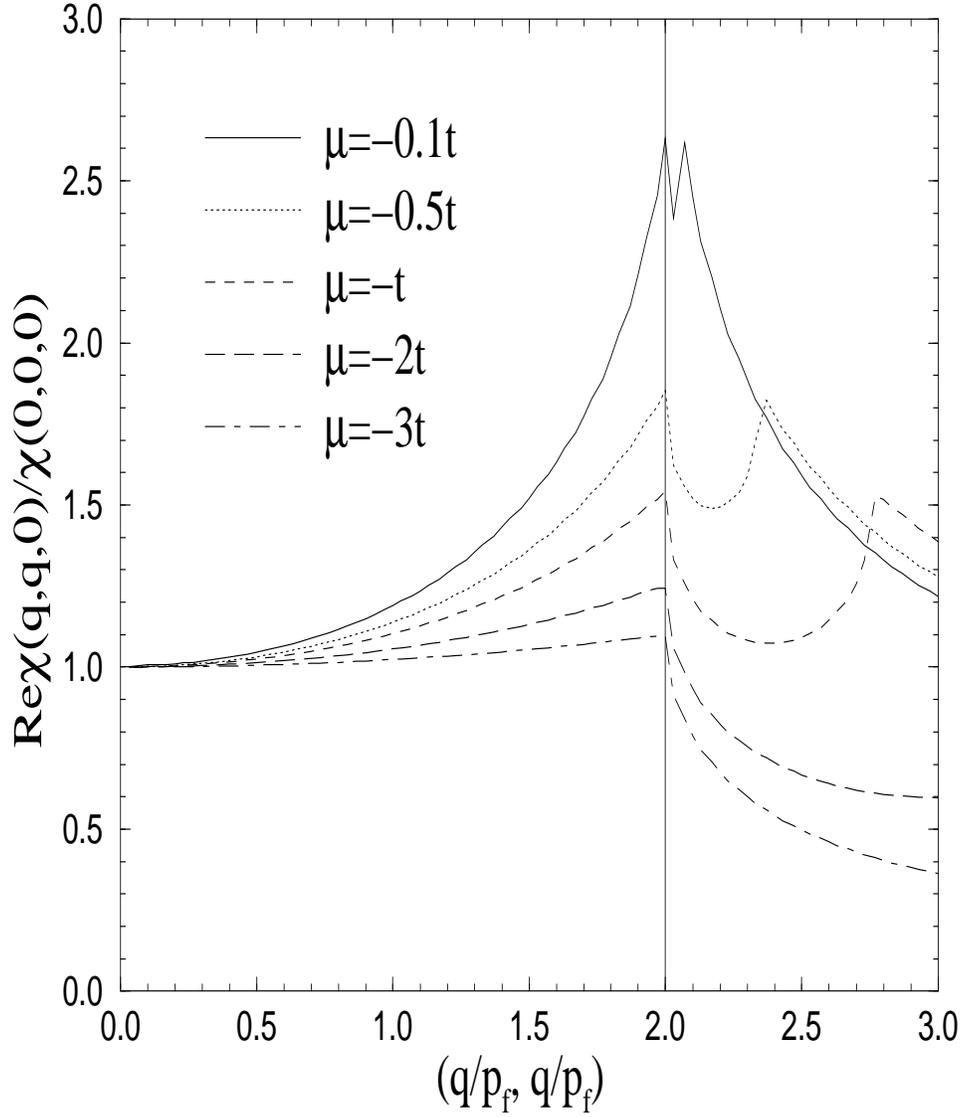,width=5in,height=6in}
\end{center}
\caption{The real part of the correlation function in static
limit. The system changes 2D to 1D as $\mu \rightarrow 0$. A
divergence grows at $q=2p_f$ as $\mu \rightarrow 0$ in $(1,1)q$
direction.}
\label{static}
\end{figure}

\begin{figure}
\begin{center}
\epsfig{file=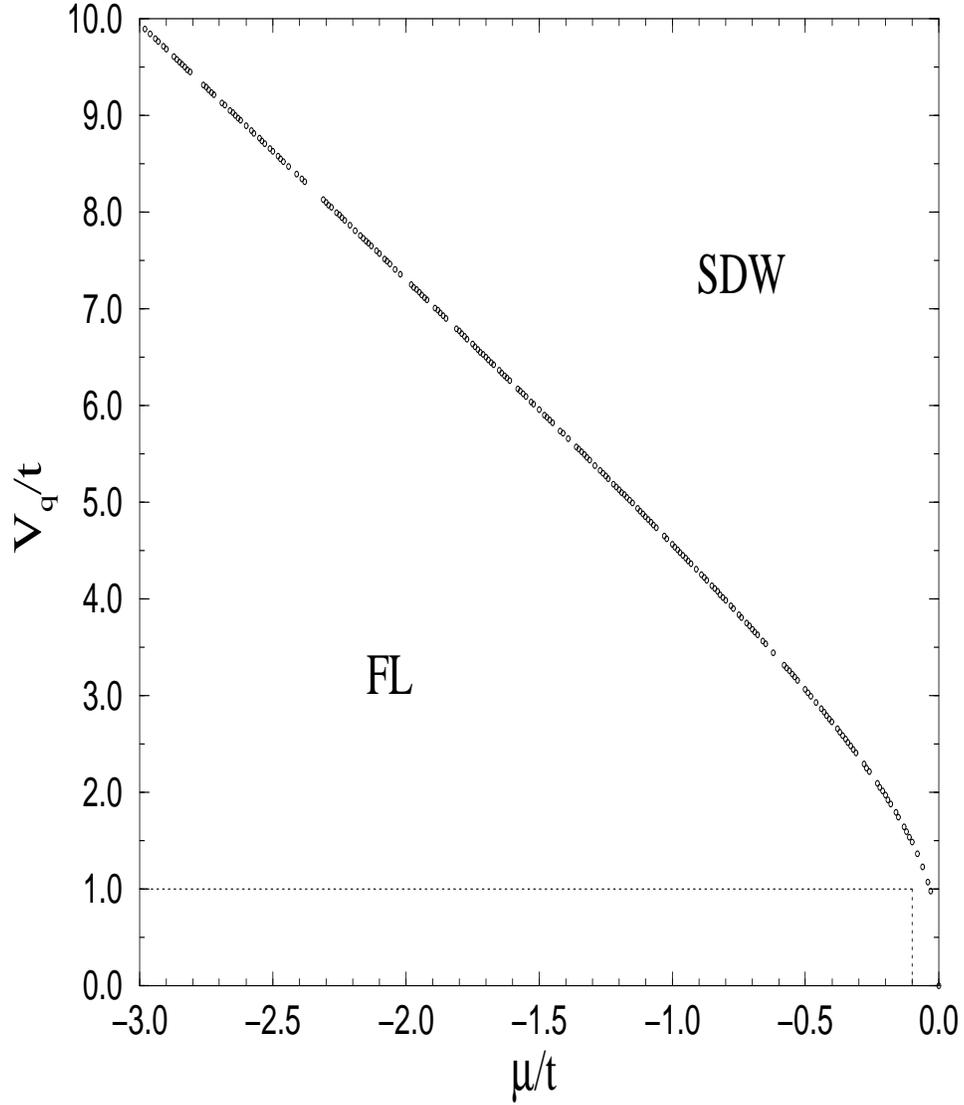,width=5in,height=6in}
\end{center}
\caption{The phase diagram of the system from
$1-V_q\chi'(2p_f,2p_f,0)=0$. For any strength of interaction at
$\mu=0$ the quasi particle picture breaks down due to the
divergence. With $V_q=t$, there is no SDW instability for $\mu <
-0.1t$.}
\label{phase}
\end{figure}

The figure \ref{static} the static correlation function in $(1,1)\vec
q$ direction has a developing divergence at $|\vec q|=2p_f^{(1,1)}$ as
the system changes 2D to 1D. The function is repeated at $|\vec
q|=\pi$ due to periodic lattice where $|\vec q|=\pi$ is at the middle
of the dip right next to the peak. $\mu =-3t$ which has the Fermi
surface close to circle shows constant $q$ dependence with densities
of state
\begin{equation}
N(\mu)=\lim_{\vec q \rightarrow 0} \lim_{\omega \rightarrow 0}
\chi'({\vec q},\omega)
\end{equation}
as in parabolic band case. The phase diagram, $V_{\vec q}$ vs. $\mu$,
is calculated by $1/V_{\vec q}=\chi'(2p_f,2p_f,\omega=0)$. With a
interaction strength of $V_q=t$, the Stoner instability does not occur
until $\mu \sim -0.03t$. In the calculations discussed in this paper
$V_q=t$ dotted line in figure \ref{phase} so that the instability of
the system is absent except when $\mu \sim 0$ for $V_q \leq t$.

This conclusion is supported by a recent renormalization group
analysis of the Hubbard model by Halboth and Metzner\cite{halboth} who
found that the tendency towards the antiferromagnetic nesting
instability is limited to $\mu \lesssim -0.1t$ for $U=t$ as evidenced
by a growing spin susceptibility. They found in fact that
$d_{x^2-y^2}$ superconductivity fluctuations provide the dominant
susceptibility further from half filling but even these were
suppressed beyond $\mu \sim -0.01t$. The influence of superconducting
fluctuations are missing in the present calculation\cite{zanchi}.

Going beyond this weak-coupling region is difficult because of the
uncontrolled nature of the RPA. One such attempt is the so-called
self-consistent fluctuation exchange approximation (FLEX) which
however is equally uncontrolled. It has been pointed out by Trembley
and coworkers that the absence of vertex corrections in the FLEX leads
even more severe breakdowns of sum rules than does the RPA discussed
here. Vilk and Trembley have proposed an improvement in RPA to ensure
that the sum rules are satisfied. In the approach the influence of the
cross channels in the $ph$ expansion on the vertices is mimicked by
letting the irreducible vertices in the charge and spin channels be
independent of each and chosen to satisfy sum rules. In this approach
SDW fluctuations above $T=0$ lead to the breakdown of the
quasi-particle approximation for $U=4t$. These approaches have been
limited to finite size systems and so miss the long wavelength
interactions responsible for the leading corrections of interest in
the present work.
\subsubsection{Contributions from the Particle-Particle Channel}
Fukuyama et al.\cite{fukuyama} have investigated the contribution to
$\Sigma''({\vec p},E)$ for the $pp$ channel using the parabolic
dispersion. The contribution to $\Sigma({\vec p},E)$ in second order
in the interaction can be thought of as being in either the $ph$ or
$pp$ channel. In the second order calculation the difference between
the two channels is that $\xi_{\vec p}^2 \ln \xi_{\vec p}$
contribution comes from propagating particle or hole pairs with
momenta $q \sim 2p_f$ in $pp$ channel whereas it comes from
longwavelength particle-hole pairs in $ph$ channel as we shown above
in figure \ref{selfdemo}. Taking the second order diagram to be in the
$pp$ channel contribution, its contribution is screened by repeated
scattering in that channel. In contrast to the case of an atractive
interaction there is no Cooper instability but instead the
contribution is reduced by the repeated scattering. Unlike the
situation in the $ph$ channel the result is not simply an enhancement
or reduction of the contribution from the second order diagram giving
$\xi_p^2 \ln \xi_p$ contribution to $\Sigma''(p,\xi_p)$ again. Rather
there is a strong momentum dependence in real part of the $pp$
propagator at $q \sim 2p_f$ which eliminates the $\xi_p^2 \ln \xi_p$
dependence. As a result the effect of including the $pp$ channel in a
calculation of $\Sigma(p,E)$ is to remove the contribution of the
second order term in the interaction from leading dependences. The
$pp$ contribution to $\Sigma(p,E)$ in a tight binding bandstructure
remains to be worked out in detail. However the result is unlikely to
be quantitatively different from the result for the parabolic
band. Indeed for particle-pairs/hole-pairs with net momentum ${\vec
Q}=(\pi, \pi)$ it is easy to show that the propagator $K({\vec Q},
\omega)$ behaviors as $K({\vec Q}, \omega) \sim {1 \over \omega + 2
\mu }f( \mu )$ where $f(\mu )$ is a polynomials in $\mu$ and gives a
vanishingly small contribution to $\Sigma(p, E)$ where the
consequences of the almost nested Fermi surface would be expected to
be greatest.

There have been a number of investigations of collective modes in the
$pp$ channel. Engelbrecht and Randeria\cite{engelbrecht} found a
collective mode below the two particle continuum in a bound state of
hole pairs which arises because of the finite density of states at
bottom of the band in 2D. The contribution of this mode to the
self-energy is given by $\Sigma''(p_f,E) \sim \mu |{E \over
E_a}|^{5/2}$ where $\mu$ is the chemical potential and $E_a$ is an
energy scale describling the interaction. So this mode does not
contribute to the leading dependences in
$\Sigma''(p,E)$. Yang\cite{yang} and more recently Demler and
Zheng\cite{demler} have found evidence for a different singlet and
triplet collective mode in the $pp$ channel which exists in a narrow
region of momenta near $(\pi, \pi)$. The narrow range of momenta
suggests that this mode can have little effect on the low energy
properties of $\Sigma(p,E)$.

In summary neither the continuum nor collective mode contributions
from the $pp$ channel contribute to the leading dependences in
$\Sigma(p,E)$.
\subsection{Off-Shell Calculations, $\Sigma({\vec p},E)$: Spectral Function}
\begin{figure}
\begin{center}
\epsfig{file=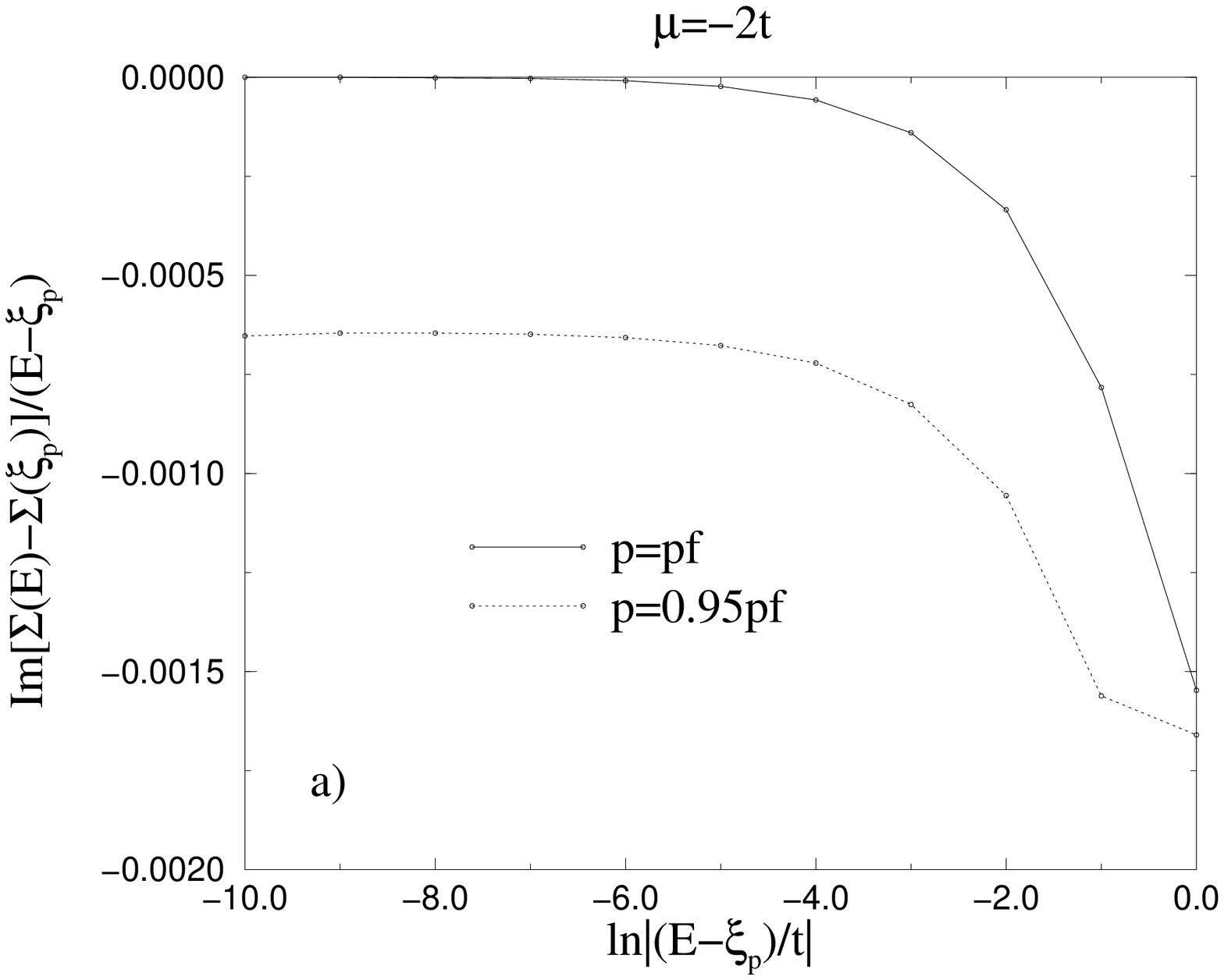,width=5in,height=3.5in}
\epsfig{file=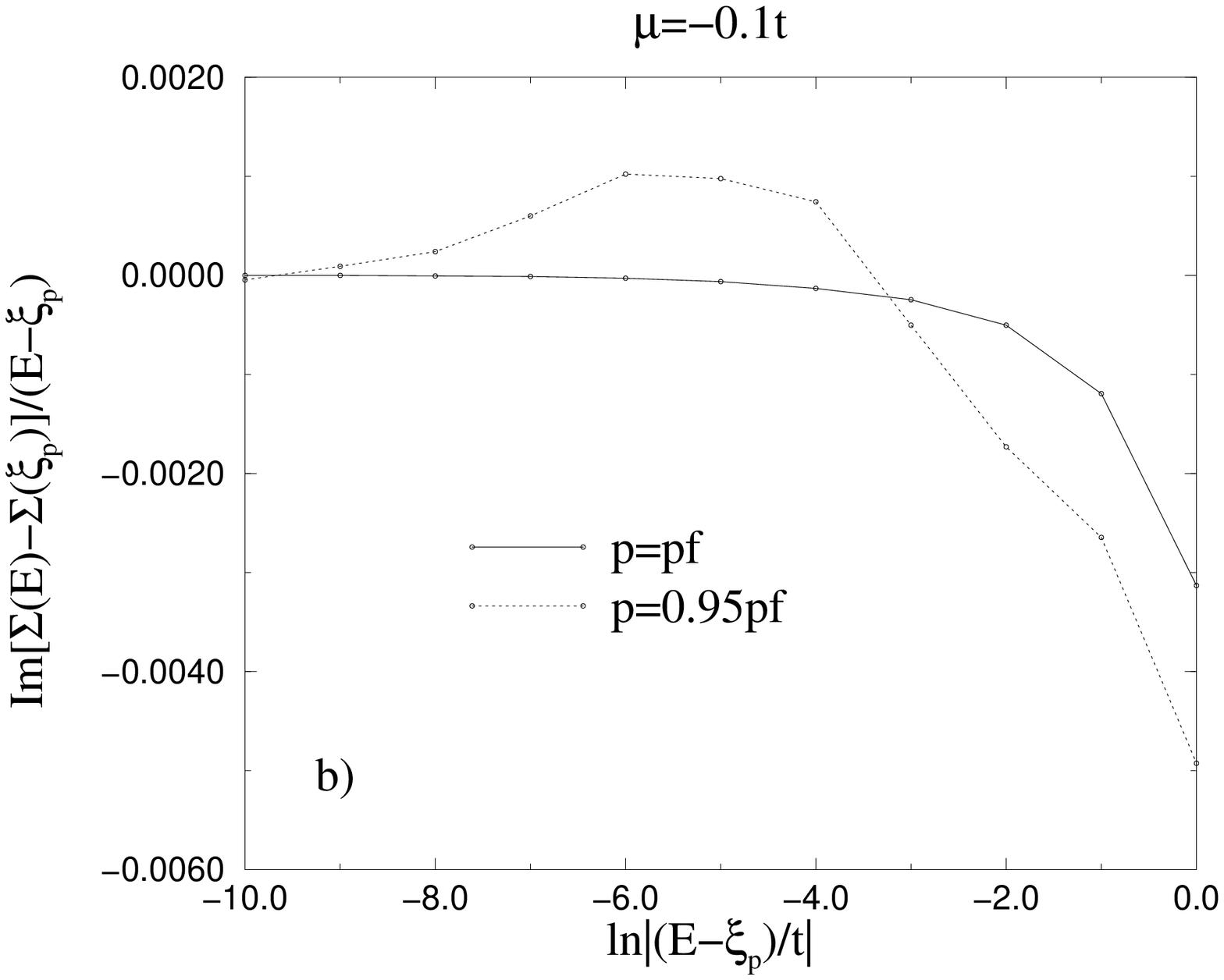,width=5in,height=3.5in}
\end{center}
\caption{The imaginary part of the off-shell self-energy for $E<0$ and
$p<p_f$ in second order calculations. $E \rightarrow \xi_{\vec p}$
limit $\Sigma''({\vec p},E)$ has constant energy term, $\xi_{\vec
p}^2\ln\xi_{\vec p}$, for $\mu=-2t$ in $ln|E-\xi_{\vec p}|$ scale in
the case of $p\neq p_f$. There is a significant deviations in
$\Sigma''({\vec p},E)$ for $\mu=-0.1t$ in the case of $p\neq p_f$.}
\label{offsefit}
\end{figure}

\begin{figure}
\begin{center}
\epsfig{file=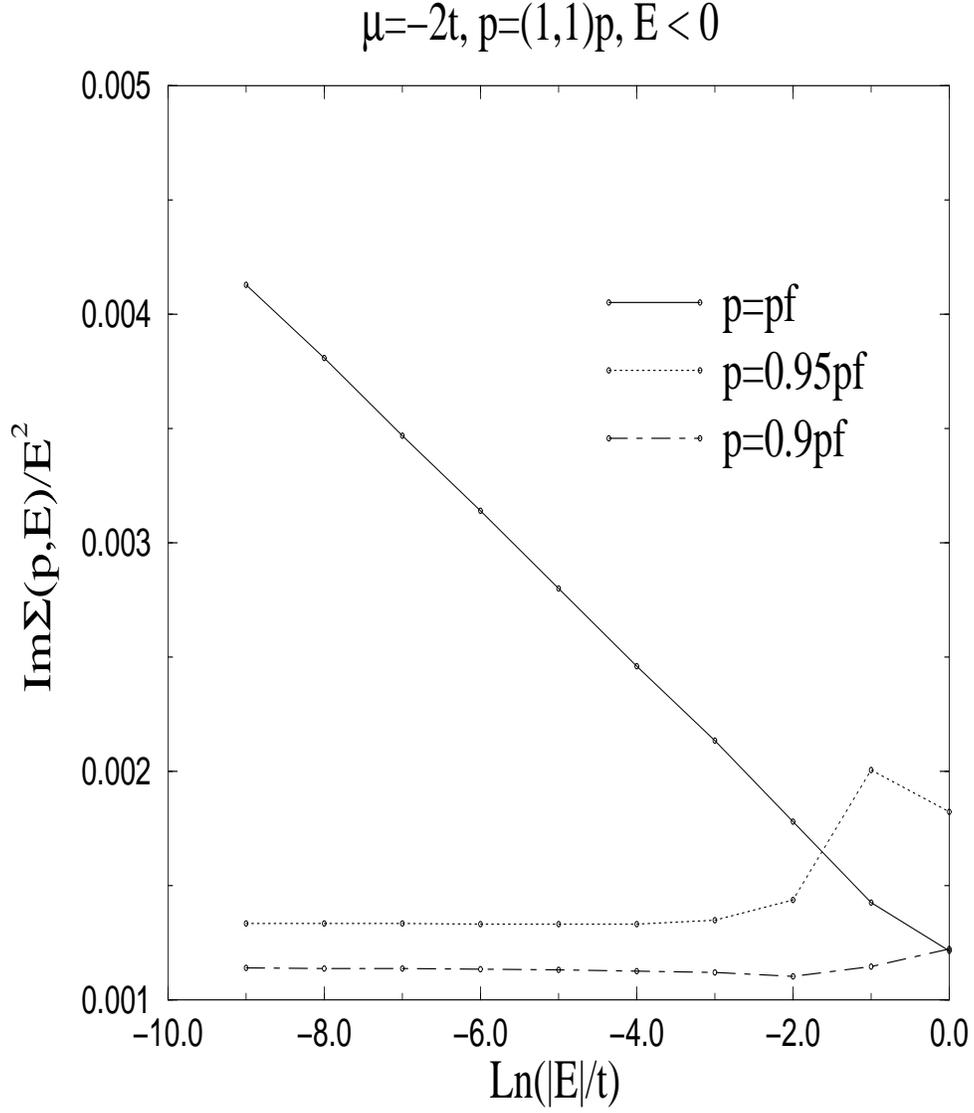,width=5in,height=6in}
\end{center}
\caption{The graph shows that the off-shell self energy follows $E^2
\ln E$ only when $p=p_f$ even for $\mu = -2t$ which is close to
parabolic band structure.}
\label{offselog}
\end{figure}

The spectral function can be measured by ARPES and is given by
\begin{equation}
A({\vec p}, E)={|\Sigma''({\vec p}, E)| \over [E - \xi_{\vec p}
-\Sigma'({\vec p}, E)]^2 +[\Sigma''({\vec p}, E)]^2}
\end{equation}
where the off-shell self-energies are involved. For large cut-off
momentum, $q_c$, and parabolic band structure the leading
contributions to the imaginary part of off-shell self-energy which
determines the width of the spectral function has a form\cite{coffey}
\begin{equation}
\Sigma''({\vec p},E)= C[\{\xi_{\vec p}^2+2\xi_{\vec p}(E-\xi_{\vec
p})\} \ln(max[\xi_{\vec p},|E|]) +(E-\xi_{\vec p})^2\ln(|E-\xi_{\vec
p}|)]
\end{equation}
where $C$ is a constant. In the figure \ref{offsefit} $[\Sigma''({\vec
p},E)-\Sigma''(\xi_{{\vec p}, \vec p})]/(E-\xi_{\vec p})$
vs. $\ln|E-\xi_{\vec p}|$ is plotted for $E-\xi_{\vec p} < 0$. In the
limit $(E-\xi_{\vec p}) \rightarrow 0$ we find $2C\xi_{\vec
p}\ln|\xi_{\vec p}|$ consistent with equation (15) for given $\vec
p$. For $\mu=-2t$ the Fermi surface is close to parabolic band
structure and the off-shell self-energy has a $(E-\xi_{\vec p})$
independent behavior for small values of $(E-\xi_{\vec p})$ as in the
equation (15). For $\mu=-0.1t$ the $\Sigma''(p_f,E)$ is consistent
with the equation (15). However for $p=0.95p_f$ although
$\Sigma''({\vec p},\xi_{\vec p})$ still has the $\xi_{\vec p}^2 \ln
\xi_{\vec p}$ dependence, $\Sigma''({\vec p},E)$ is seen to deviate
from the form given in equation (15). The dependence on $E$ in figure
\ref{offselog} shows a $E^2\ln |E|$ behavior for $\mu=-2t$ only for $p
= p_f$ and the more complicated dependence shown in equation (15) is
found for $|{\vec p}| \ne p_f$. This is the case for other values of
$\mu$ also. This indicates that the form $E^2 \ln E$ for $\Sigma({\vec
p},E)$ frequently used in the literature is a poor
approximation\cite{jackeli,menashe} to the more complicated function
of $\vec p$ and $E$.

ARPES data is analyzed in terms of the single-particle spectral
density. The extent to which a quasi-particle resonance is well
defined is characterized by the width of the resonance at
half-maximum. This is frequently thought of as a measure of the
quasi-particle life-time where momentum dependence is taken as a test
of the Fermi liquid character of the material under
investigation. When the quasi-particle approximation is applied, the
spectral function is
\begin{equation}
A({\vec p}, E)={\Gamma_{\vec p} \over (E-E_{\vec p})^2 +\Gamma_{\vec
p}^2}
\end{equation}
where $E_{\vec p}$ is the location of the resonance and $\Gamma_{\vec
p}$ is half-maximum width of the peak. Here we apply this analysis to
the Hubbard model.

\begin{figure}
\begin{center}
\epsfig{file=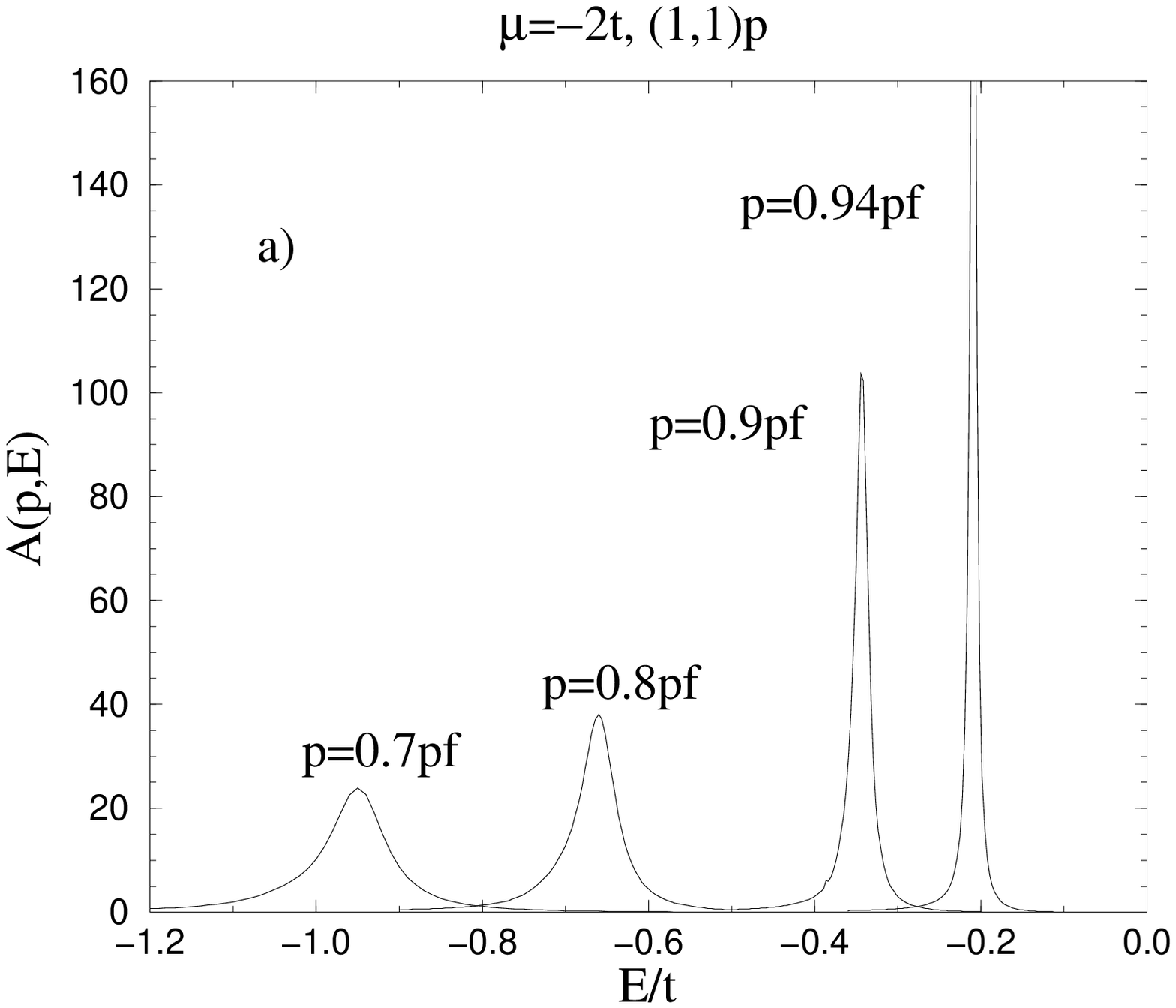,width=5in,height=4in}
\epsfig{file=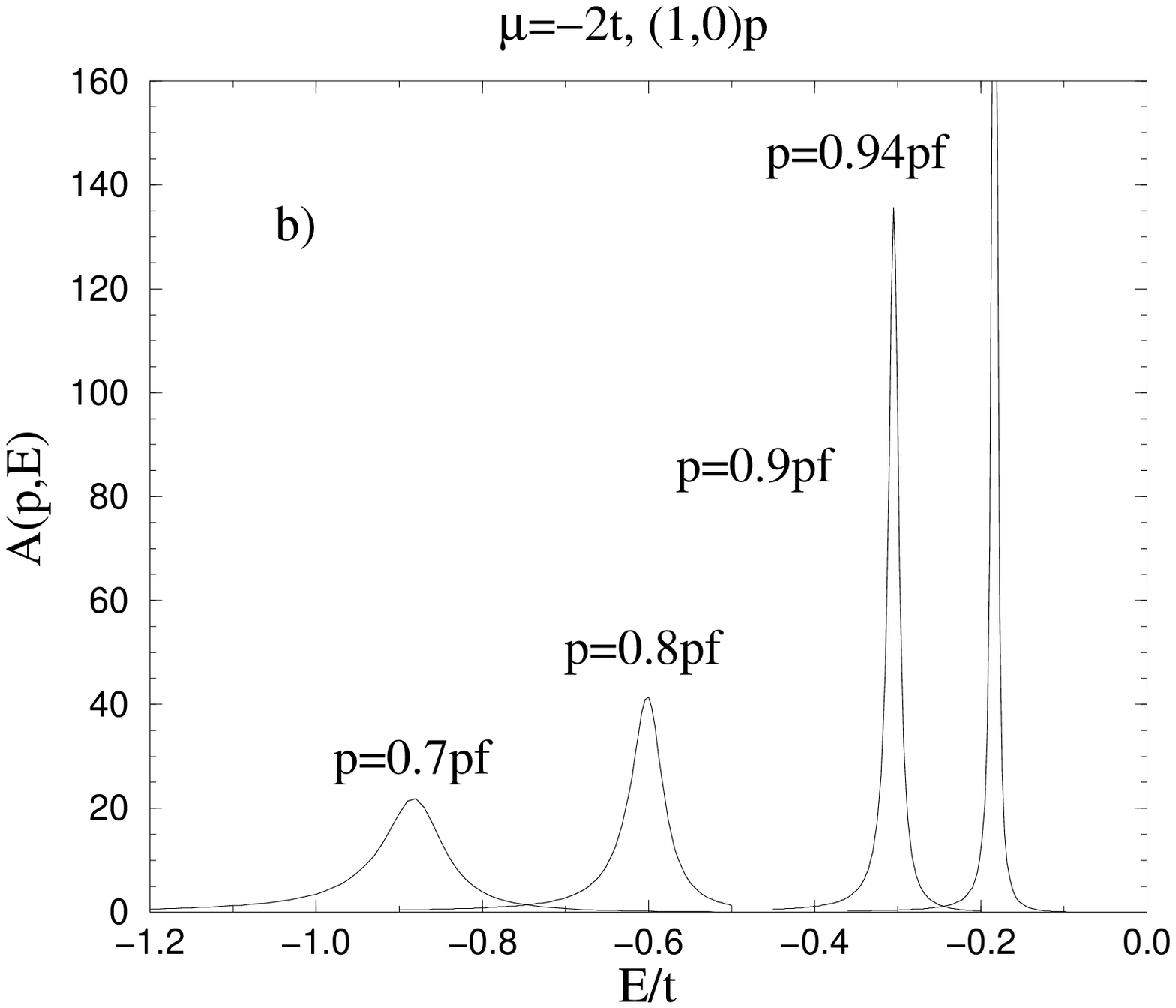,width=5in,height=4in}
\end{center}
\caption{The spectral functions in repeated scattering calculations
for $\mu=-2t$ in $(1,1)\vec p$ (a) and $(1,0)\vec p$ (b)
directions. The quasi-particle peakes are well defined as $p
\rightarrow p_f$. Asymmetry in the resonance peak apears when $p \sim
p_f$.}
\label{arpes2.0}
\end{figure}

\begin{figure}
\begin{center}
\epsfig{file=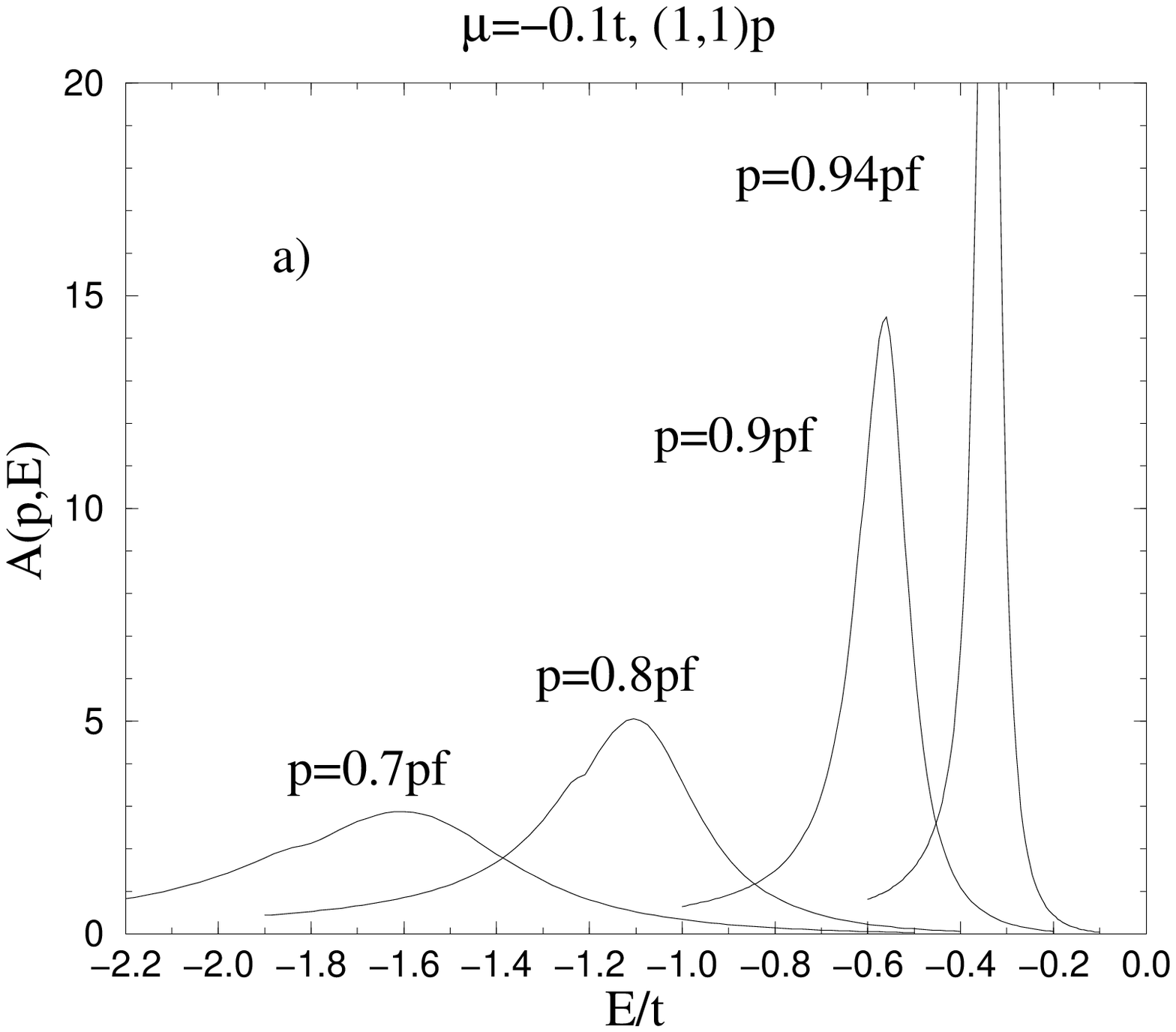,width=5in,height=4in}
\epsfig{file=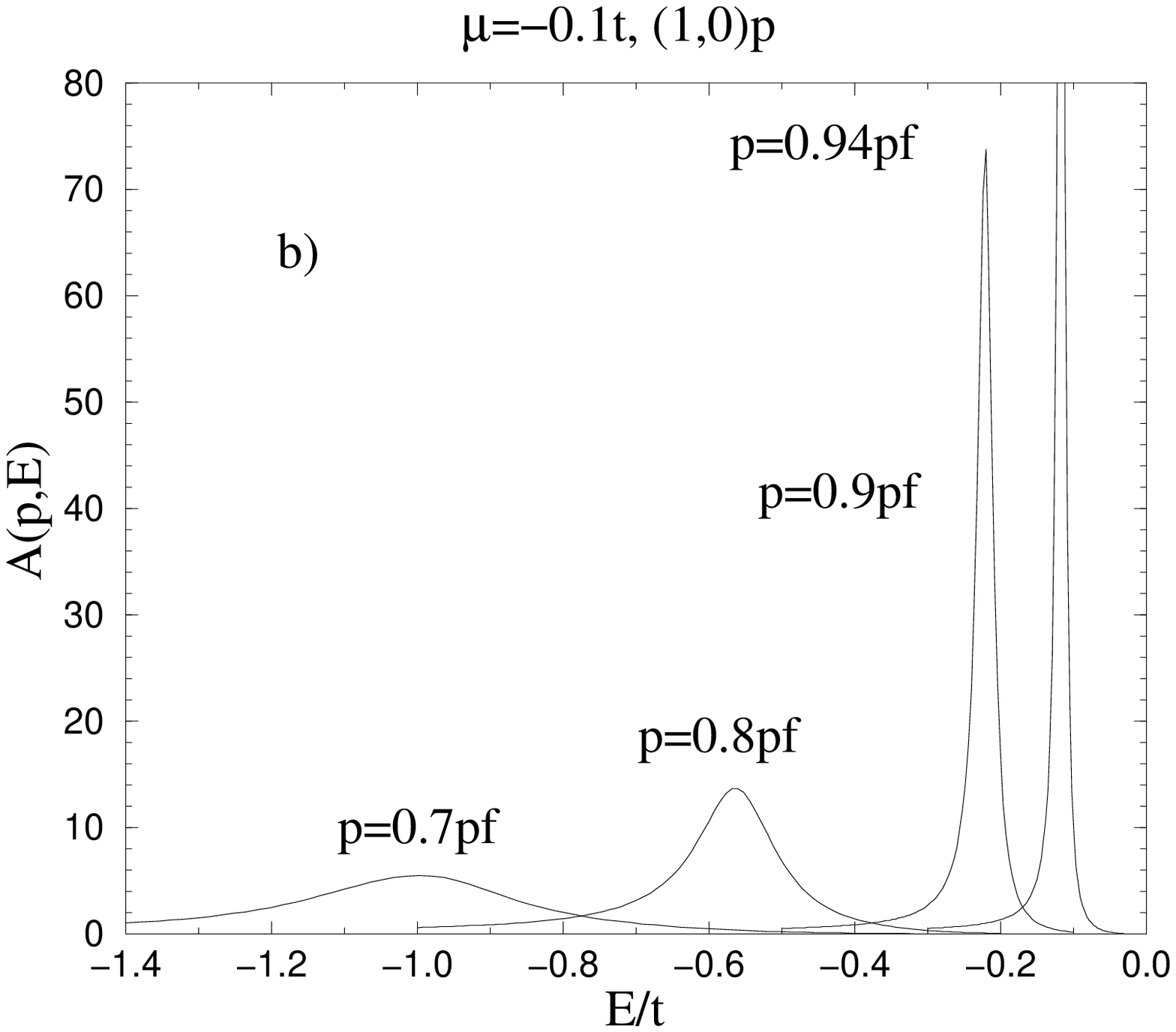,width=5in,height=4in}
\end{center}
\caption{The spectral functions in repeated scattering calculations
for $\mu=-0.1t$ in $(1,1)\vec p$ (a) and $(1,0)\vec p$ (b)
directions. The quasi-particle peakes are well defined as $p
\rightarrow p_f$. Asymmetry in the resonance peak apears for
$(1,1)\vec p$ in deep inside of Fermi surface.}
\label{arpes0.1}
\end{figure}

The spectral functions are calculated for $\mu=-2t$ and $\mu=-0.1t$
cases and the widths of the resonance at the half of the peak,
$\Gamma_{\vec p}$, are plotted as a function of $\xi_{\vec p} <
0$. The spectral functions are picked along $\Gamma$ to $Y$ point and
to $\bar M$ point. The quasi-particle peaks are well defined and as $p
\rightarrow p_f$ resonance gets sharper for $(1,0)p$ and $(1,1)p$
directions in the figure \ref{arpes2.0} for $\mu=-2t$. The $\mu=-0.1t$
case in the figure \ref{arpes0.1} has the same feature except growing
anisotropic feature for $(1,1)p$ direction in deep inside of the Fermi
surface. There is no sign for the pseudogap with the parameters used
in this calculations for $\mu=-2t$ and $\mu=-0.1t$. As $p \rightarrow
p_f$ the resonance peak does not show symmetry about $E=E_{\vec p}$
since the imaginary part of self energy is always zero for
$E=0$. $A({\vec p},E)$ is only symmetric about $E=0$ for $|{\vec
p}|=|p_f|$. The smallest deviation from $p_f$ leads to strongly
non-symmetric $A({\vec p},E)$. This indicates that one can be misled
by assuming symmetry for the ARPES data cutoff by Fermi Dirac
distribution function\cite{norma2} to get the single-particle
properties near Fermi surface.

\begin{figure}
\begin{center}
\epsfig{file=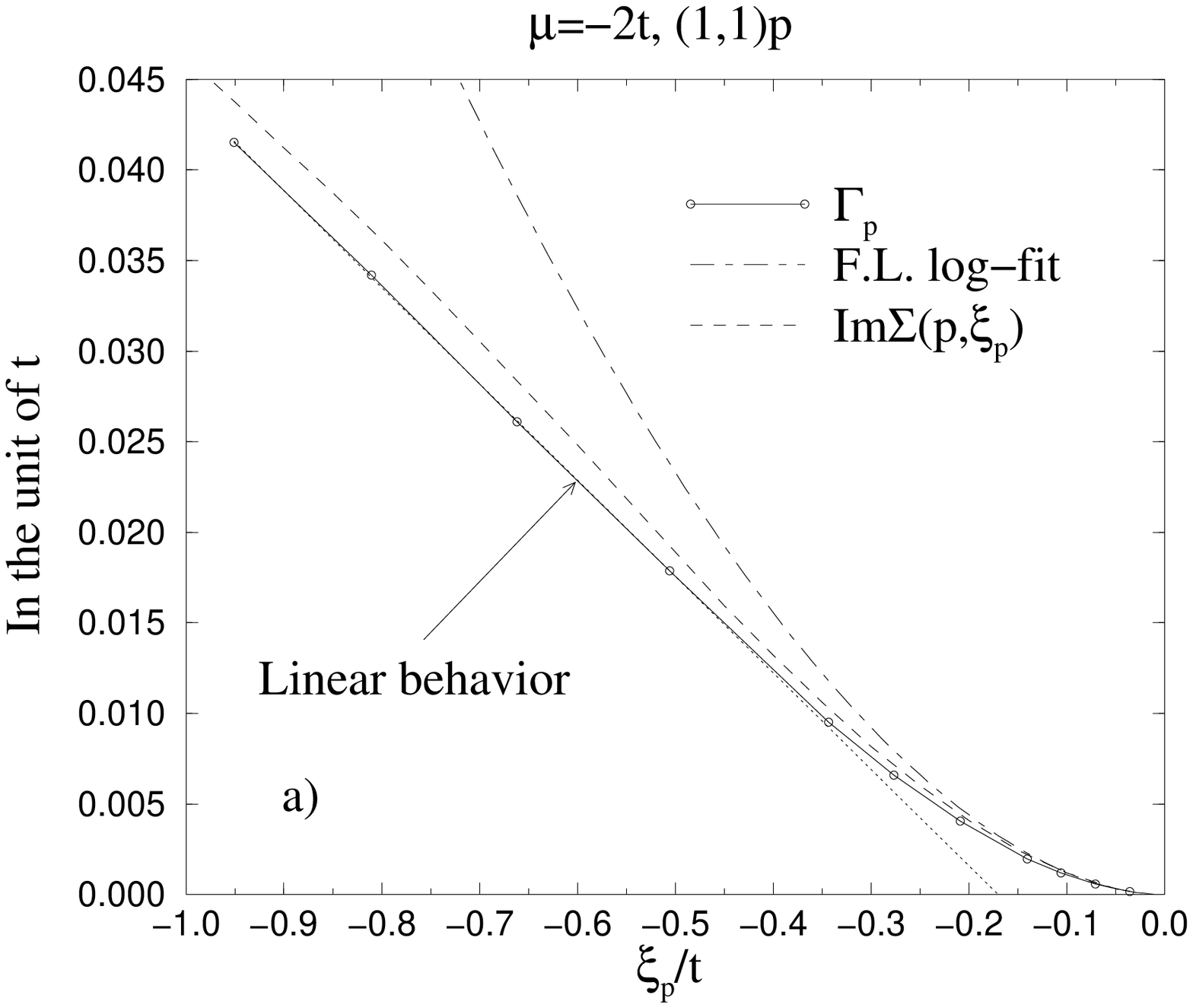,width=5in,height=4in}
\epsfig{file=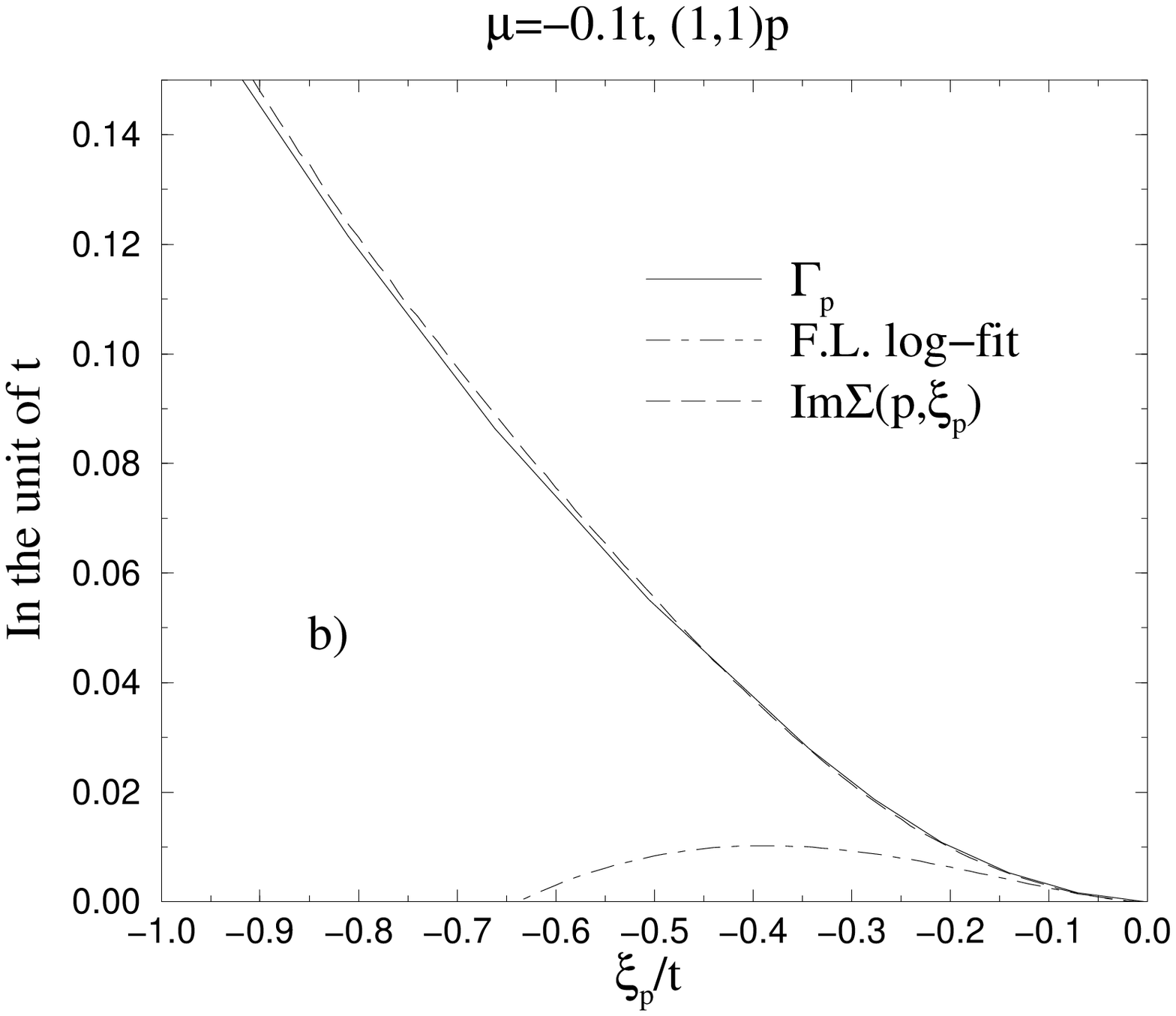,width=5in,height=4in}
\end{center}
\caption{The comparison between $\Gamma_{\vec p}$ and on-shell
self-energy for $\mu=-2t$ (a) and $\mu=-0.1t$ (b). The FL behavior,
$\xi_{\vec p}^2 \ln \xi_{\vec p}$, is restricted to low energies.}
\label{acompare}
\end{figure}

If we expand $A({\vec p},E)$ near $E_{\vec p}=\xi_{\vec
p}+\Sigma'({\vec p},E)$ which is approximately the location of the
resonance peak since $\Sigma({\vec p},E)$ is continuous, the width of
half maximum is given by the on-shell self-energy
\begin{equation}
\Gamma_{\vec p} \sim {z_{\vec p}\Sigma''({\vec p},E_{\vec p}) \over
\sqrt{1+[ z_{\vec p}\partial \Sigma''({\vec p},E) / \partial E
|_{E=E_{\vec p}}]^2}}
\end{equation}
where $z_{\vec p}=[1- \partial \Sigma'({\vec p},E) / \partial E
|_{E=E_{\vec p}}]^{-1}$ unless $\Sigma''({\vec p},E)$ has a strong
energy dependences. In this calculation we find $1-z_{\vec p} \lesssim
3 \%$ showing that there is very little frequency dependence in the
real part, $\Sigma'({\vec p},E) \sim \Sigma'({\vec p}, \xi_{\vec
p})$. The figure \ref{acompare} shows the agreement between the half
maximum width and the on-shell self-energy at low energies,
$\Gamma_{\vec p} \sim \Sigma''({\vec p}, \xi_{\vec p})$, for $\mu=-2t$
and $\mu=-0.1t$. As in the figure the generic FL logarithmic behavior
is restricted to low energies. For $\mu=-2t$ case $\Gamma_{\vec p}$
shows linear behavior in $\xi_{\vec p}$ at higher energies which is
not extrapolated to zero energy. Clearly the assumption that a
deviation from $\Gamma_{\vec p} \sim \xi_{\vec p}^2 \ln \xi_{\vec p}$
or $\xi_{\vec p}^2$ is an indication of non-Fermi liquid is not
correct. We note that no sign of the complicated $\xi_{\vec p}$ and
$E$ dependences of $\Sigma({\vec p}, E)$, shown in equation (15),
comes out of this analysis of ARPES data. It seems to us unlikely that
it will be possible to recover these dependences without prior
knowledge of the functional form to which to fit the data.
\section{Conclusion and Summary}
We have demonstrated that anisotropy in band structure does not
qualitatively change FL dependencies. We have done that by calculating
single-particle properties in the Hubbard model in which the shape of
the Fermi surface is changed from a circle to a square as the value of
the chemical potential is varied from $\mu=-3t$ to $\mu=-0.1t$. In
these calculations we have used a weak coupling approximation in which
the magnitude of the interaction, $V_q$, is far from the values for
the SDW instability except for $\mu > -0.5t$. Numerical coefficients
in the functional forms of the imaginary part of the self-energy
reflect the anisotropy of the band structure but it is not until the
Fermi surface is almost a square that the first sign of quasi-1D
behavior, $\xi_{\vec p}^{3/2}$, appears. As $\mu \rightarrow 0$ the
magnitude of $\Sigma''({\vec p},\xi_{\vec p})$ increases with with
increasing anisotropy. Also, in the $\mu \rightarrow 0$ limit, an
$\omega$ independent region in $\chi''({\vec q},\omega)$ grows, which
is responsible for $\xi_{\vec p}^{3/2}$ behavior in $\Sigma''({\vec
p},\xi_{\vec p})$ in $(1,0)p$ direction. The $\xi_{\vec p}|\xi_{\vec
p}|$ term is also found in $\Sigma'({\vec p},\xi_{\vec p})$ as in
parabolic band case. The instability of the system is slowly developed
as the system goes to 1D except $\mu \sim 0$.

The off-shell self-energy is shown to have a similar $\vec p$ and $E$
dependence as in the parabolic band with deviations as the Fermi
surface developes flat regions. There is no evidence for the pseudogap
phenomenon for the parameters used here, $V=t$ and $\mu < -0.1t$. In
this region we confirmed the half maximum width of the resonance peak
for spectral function is approxmated by on-shell
self-energy. Asymmetry in the resonance peak grows for $|{\vec
p}|=p_f$ due to $\Sigma''({\vec p},E=0)=0$ and for deep inside of
Fermi surface for $\mu=-0.1t$ and $(1,1)|\vec p|$ direction. In the
tight binding band structure quasi-particle picture is well defined
for $\mu < -0.1t$ and $|V_{\vec q} \lesssim t$.

As one goes further away from this parameter region it becomes
difficult to justify the simple RPA weak-coupling approximation used
here. SDW and superconducting fluctuations are likely to play a role
as other authors have pointed out. Their effects on the leading
dependences from long wavelengths in $\Sigma({\vec p},E)$ remain to be
investigated.


\begin{figure}
\begin{center}
\epsfig{file=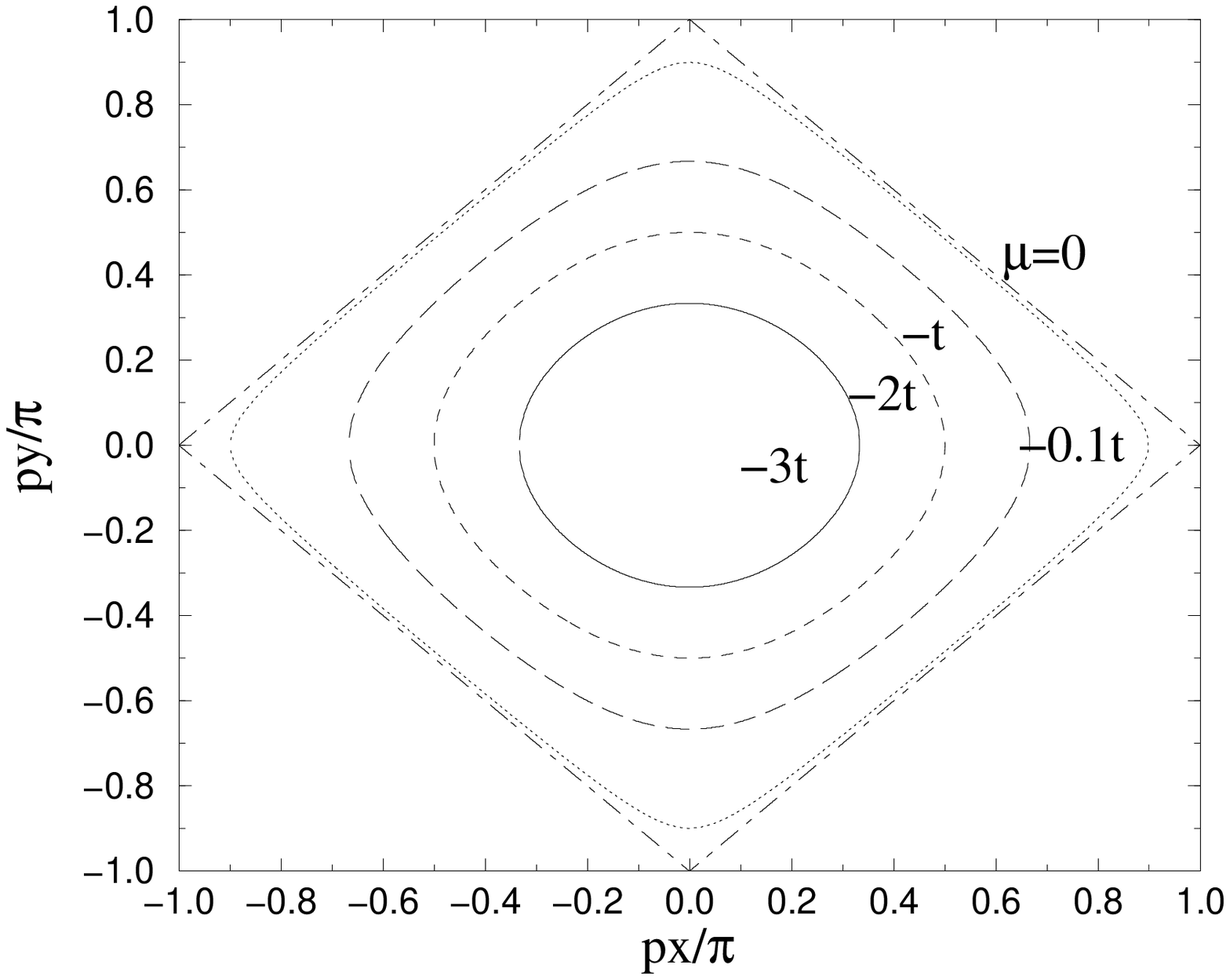,width=5.5in,height=6in}
\end{center}
\caption{The band structures for the tight-binding dispersion,
$\xi_{\vec p}=-2t\cos p_x-2t\cos p_y -\mu$ with different values of
$\mu$. As band is filled up, the parabolic Fermi surface is changes
toa square surface. For $\mu =-0.1t$, there is a large region of flat
Fermi surface.}
\label{band}
\end{figure}

\begin{figure}
\begin{center}
\epsfig{file=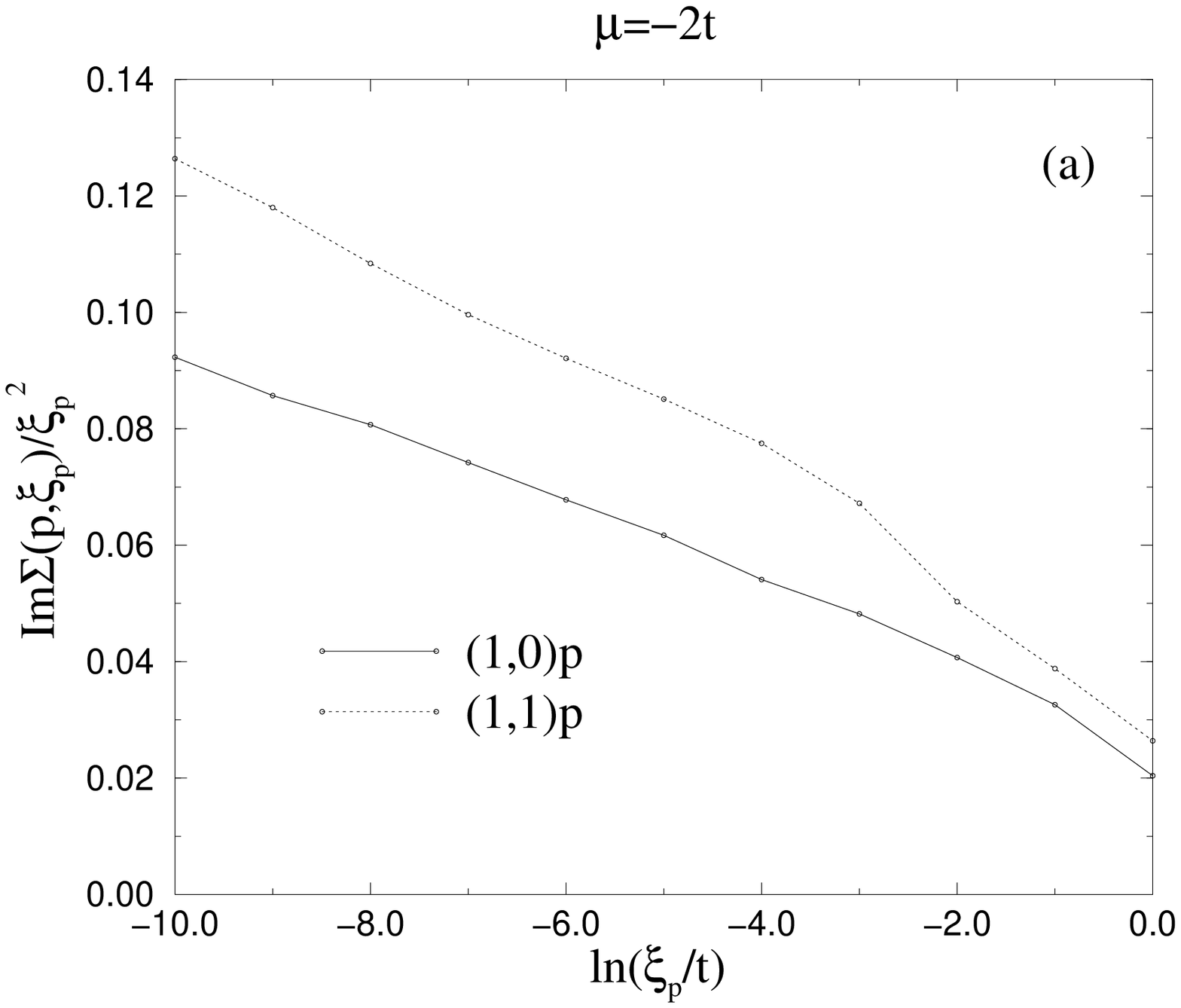,width=4in,height=4in}
\epsfig{file=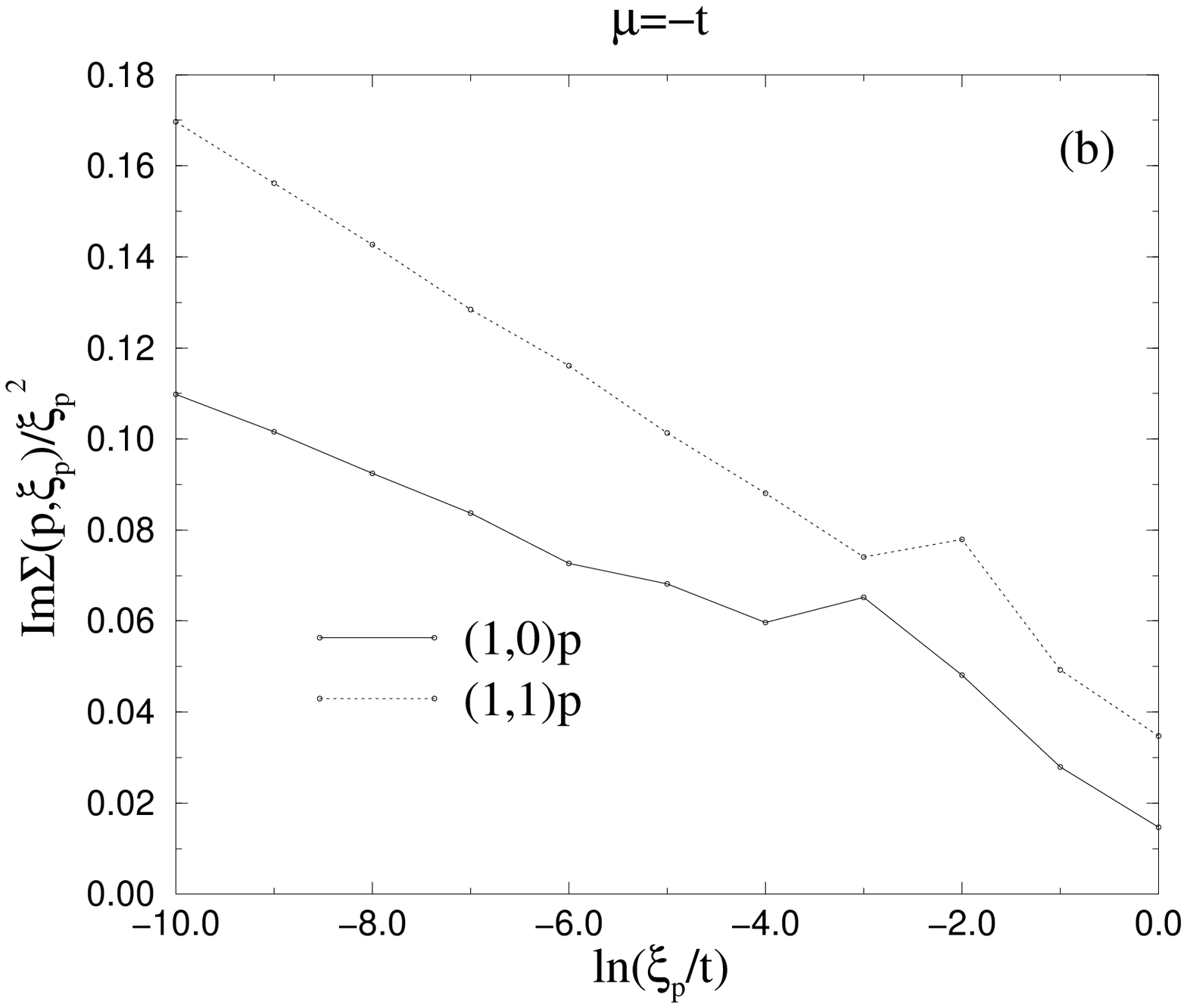,width=4in,height=4in}
\end{center}
\caption{The imaginary part of self-energies in second order
calculations for $(1,0)p$ and $(1,1)p$ directions with (a) $\mu=-2t$
and (b) $\mu=-t$. Anisotropy increases in the magnitude of the
self-energy as $\mu$ increases.}
\label{2ndse}
\end{figure}

\begin{figure}
\begin{center}
\epsfig{file=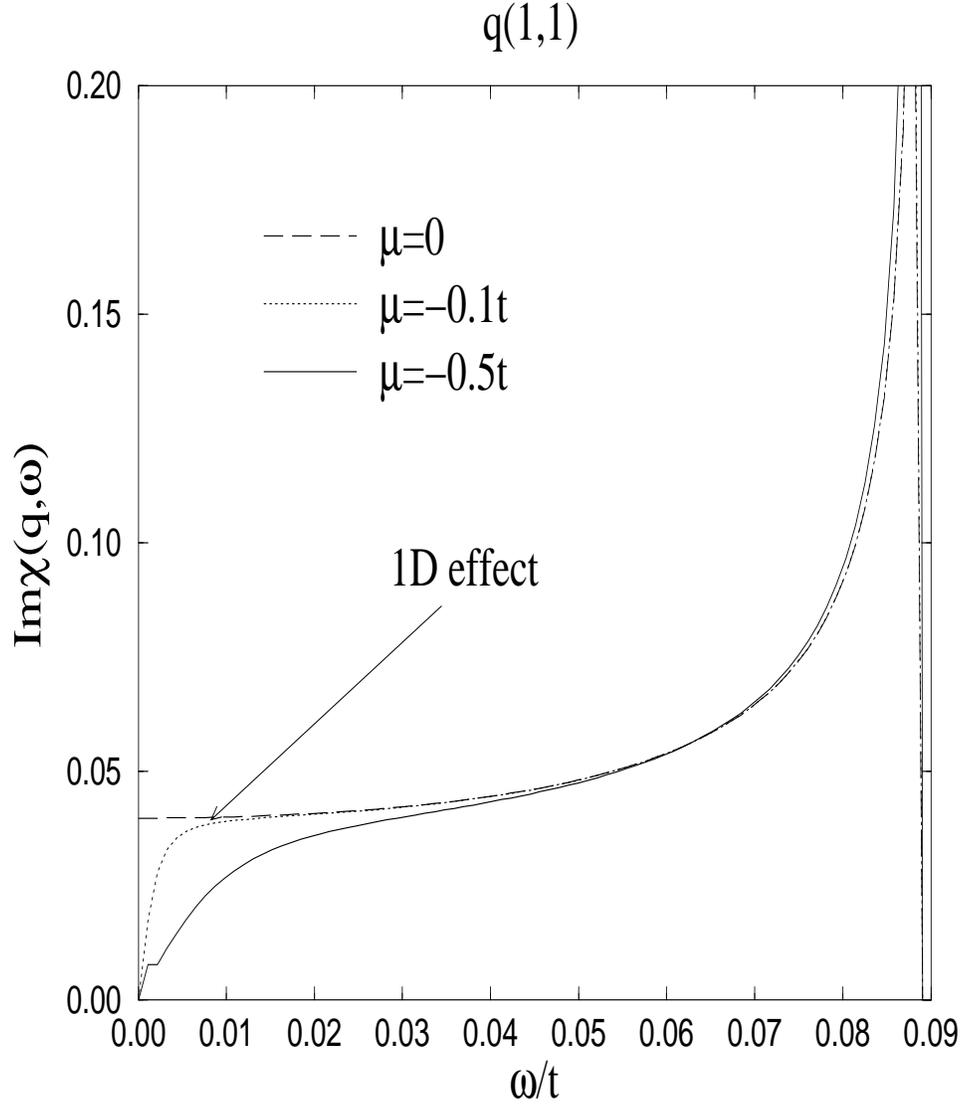,width=5in,height=6in}
\end{center}
\caption{The correlation functions for $q=0.01\pi$ with different
$\mu$s. As $\mu \rightarrow 0$ $\omega$ independent behavior in low
frequency is develops which is characteristic of 1D feature at long
wavelengths and the system shows 1D effect.}
\label{chilong}
\end{figure}

\begin{references}
\bibitem{pines}D. Pines and P. Nozieres, {\it The Theory of Quantum
Liquids}, Vol.1. Benjamin, New York (1966).
\bibitem{bednorz}J. G. Bednorz and K. A. M\"{u}ller, Z. Phy. B {\bf
64}, 189 (1986).
\bibitem{timusk}T. Timusk and R. B. Tanner, in {\it Infrared
Properties of High $T_c$ Superconductors}, Vol.1({\it Ed.}
D. M. Ginsberg), (World Scientific, Singapore, 1988).
\bibitem{takagi}H. Takagi, B. Batlogg, H. L. Kao, J. Kwo, R. J. Cava,
J. J. Krajewski and W. F. Peck, Phys. Rev. Lett. {\bf 69}, 2975
(1992).
\bibitem{ando1}Y. Ando, G. S. Boebinger, A. Passner, T. Kimura and
K. Kishio, Phys. Rev. Lett. {\bf 75}, 4662 (1995).
\bibitem{ando2}Y. Ando, G. S. Boebinger, A. Passner, N. L. Wang,
C. Geibel and F. Steglich, Phys. Rev. Lett. {\bf 77}, 2065 (1996).
\bibitem{schabel}M. C. Schabel, C. H. Park, A. Matsuura, Z. X. Shen,
D. A. Bonn, R. Liang and W. N. Hardy, Phys. Rev. B. {\bf 57}, 6090 and
6107 (1998).
\bibitem{norman} M. R. Norman, H. Ding, M. Randeria, J. C. Campuzano,
T. Yokoya, T. Takeuchi, T. Takahashi, T. Mochiku, K. Kadowaki,
P. Guptasarma and D. G. Hinks, Nature {\bf 392}, 157 (1998); J. Mesot,
A. Kaminski, H. M. Fretwell, M. Randeria, J. C. Campuzano, H. Ding,
M. R. Norman, T. Takeuchi, T. Sato, T. Yokoya, T. Takahashi, I. Chong,
T. Terashima, M. Takano, T. Mochiku, K. Kadowaki, cond-mat 9910430
\bibitem{anderson}P. W. Anderson, Phys. Rev. Lett. {\bf 64}, 1839
(1990) and {\bf 65}, 2306 (1990).
\bibitem{castellani1}C. Castellani, C. Di Castro and W. Metzner,
Phys. Rev. Lett. {\bf 69}, 1703 (1992).
\bibitem{castellani2}C. Castellani, C. Di Castro and W. Metzner,
Phys. Rev. Lett. {\bf 72}, 316 (1994).
\bibitem{boies}D. Boies, C. Bourbonnais and A. -M. S. Tremblay,
Phys. Rev. Lett. {\bf 74}, 968 (1995).
\bibitem{kopietz}P. Kopietz, V. Meden and K. Sch\"{o}nhammer,
Phys. Rev. B {\bf 56}, 968 (1995).
\bibitem{ding}H. Ding, M. R. Norman. T. Yokaya, T. Takeuchi,
M. Randeria, J. C. Campuzano, T. Takahashi, T. Mochiku and
K. Kadowaki, Phys. Rev. Lett. {\bf 78}, 2628 (1997).
\bibitem{loeser}A. G. Loeser, Z. -X. Shen, M. C. Schabel, C. Kim,
M. Zhang, A. Kapitulnik and P. Fournier, Phys. Rev. B. {\bf 56}, 14185
(1997).
\bibitem{mesot}J. Mesot, A. Kaminski, H. M. Fretwell, M. Randeria,
J. C. Campuzano, H. Ding, M. R. Norman, T. Takeuchi, T. Sato,
T. Yokoya, T. Takahashi, I. Chong, T. Terashima, M. Takano,
T. Mochiku and K. Kadowaki, cond-mat 9910430.
\bibitem{trugman}S. A. Trugman, Phys. Rev. Lett. {\bf 65}, 500 (1990).
\bibitem{moreo}A. Moreo, D. J. Scalapino, R. L. Sugar, S. R. White and
N. E. Bickers, Phys. Rev. B {\bf 41}, 2313 (1990).
\bibitem{eder}R. Eder and Y. Ohta, Phys. Rev. B {\bf 51}, 6041 (1995).
\bibitem{dagotto}E. Dagotto, A. Nazarenko and M. Boninsegni,
Phys. Rev. Lett. {\bf 73}, 728 (1994).
\bibitem{kim1}J. Kim and D. Coffey, Phys. Rev. B {\bf 57}, 542 (1998).
\bibitem{hodges}C. Hodges, H. Smith and J. W. Wilkins, Phys. Rev. {\bf
4}, 302 (1971).
\bibitem{bloom}P. Bloom, Phys. Rev. B {\bf 12}, 125 (1975).
\bibitem{coffey}D. Coffey and K. Bedell, Phys. Rev. Lett. {\bf 71},
1043 (1993).
\bibitem{williams}J. M. Williams {\it et al.}, {\it Organic
Superconductors (including fullerences) : synthesis, structure,
properties and theory} (Prentice Hall, Englewood Cliffs, 1992).
\bibitem{kim4}J. Kim and D. Coffey, unpublished.
\bibitem{kim2}J. Kim and D. Coffey, Philos. Mag. B {\bf 74}, 477
(1996).
\bibitem{virosztek}A. Virosztek and J. Ruvalds, Phy. Rev. B {\bf 42},
4064 (1990).
\bibitem{ruvalds}J. Ruvalds and A. Virosztek, Phy. Rev. B {\bf 43},
5498 (1991).
\bibitem{hlubina}R. Hlubina and T. M. Rice, Phys. Rev. B {\bf 51},
9253 (1995).
\bibitem{ioffe}L. B. Ioffe and A. J. Millis, Phys. Rev. B {\bf 58},
11631 (1998).
\bibitem{furukawa1}N. Furukuyama and T. M. Rice, J. Phys. C. {\bf 10}
L381 (1998).
\bibitem{furukawa2}N. Furukuyama and T. M. Rice, Phys. Rev. Lett. {\bf
81} 3195 (1998).
\bibitem{dessau}D. S. Dessau, Z. -Z. Shen, D. M. King, D. S. Marshall,
L. W. Lombardo, P. H. Dickinson, A. G. Loeser, J. DiCarlo,
C. -H. Park, A. Kapitulnik and W. E. Spicer, Phys. Rev. Lett. {\bf
71}, 2781 (1993).
\bibitem{hirsch}J. E. Hirsch and D. J. Scalapino,
Phys. Rev. Lett. {\bf 56}, 2732 (1986).
\bibitem{lee}P. A. Lee, and N. Read, Phys. Rev. Lett. {\bf 58}, 2691
(1987).
\bibitem{rice}M. J. Rice and S. Str\"{a}ssler, Solid State
Commun. {\bf 13}, 125 (1973).
\bibitem{halboth}C. J. Halboth and W. Metzner, cond-mat 9908471.
\bibitem{zanchi}D. Zanchi and H. J. Schulz, Eur. Phys. Lett. {\bf 44}
235 (1998); cond-mat 9812303.
\bibitem{fukuyama}H. Fukuyama and Y. Hasegawa, Prog. Theor. Phys
Suppl. {\bf 101} 441 (1990); H. Fukuyama, Y. Hasegawa and O. Narikiyo,
J. of Phys. Soc. Jpn. {\bf 60}, 2013 (1991).
\bibitem{engelbrecht}J. R. Engelbrecht and M. Randeria,
Phys. Rev. Lett. {\bf 65} 1032 (1990); Phys. Rev. B {\bf 45} 12419
(1992).
\bibitem{yang}C. N. Yang, Phys. Rev. Lett. {\bf 63}, 2144 (1989).
\bibitem{demler}E. Demler and S-C. Zheng, Phys. Rev. Lett. {\bf 75},
4126 (1995).
\bibitem{jackeli}G. Jackeli and V. Y. Yushankhai, Phys. Rev. B {\bf
56}, 3540 (1997).
\bibitem{menashe}D. Menashe and B. Laikhtman, Phys. Rev. B {\bf 59}
(1999).
\bibitem{norma2}M. R. Norman, H. Ding, H. Fretwell, M. Randeria and
J. C. Campuzano, Phys. Rev. B {\bf 60}, 7585 (1999).

\end{references}
\end{document}